\documentclass[pre,eqsecnum,preprint,showkeys,preprintnumbers,amssymb,amsfonts]{revtex4}

\usepackage{amsfonts}
\usepackage{times}

\newcommand{\beq}{\begin{equation}}
\newcommand{\eeq}{\end{equation}}
\newcommand{\beqs}{\begin{eqnarray}}
\newcommand{\eeqs}{\end{eqnarray}}
\newcommand{\beqn}{\begin{eqnarray*}}
\newcommand{\eeqn}{\end{eqnarray*}}

\newtheorem{theo}{Theorem}[section]
\newtheorem{cor}{Corollary}[section]
\newtheorem{lemma}{Lemma}[section]
\newtheorem{defi}{Definition}[section]

\newtheorem{propo}{Proposition}[section]

\begin{document}

\title{Structure of spanning trees on the two-dimensional Sierpinski gasket}

\author{Shu-Chiuan Chang$^a$} 
\email{scchang@mail.ncku.edu.tw} 

\author{Lung-Chi Chen$^b$} 
\email{lcchen@math.fju.edu.tw}
\affiliation{(a) \ Department of Physics \\
National Cheng Kung University \\
Tainan 70101, Taiwan} 

\affiliation{(b) \ Department of Mathematics \\
Fu Jen Catholic University \\
Taipei 24205, Taiwan}

\begin{abstract}
Consider spanning trees on the two-dimensional Sierpinski gasket $SG(n)$ where stage $n$ is a non-negative integer. For any given vertex $x$ of $SG(n)$, we derive rigorously the probability distribution of the degree $j \in \{1,2,3,4\}$ at the vertex and its value in the infinite $n$ limit. Adding up such probabilities of all the vertices divided by the number of vertices, we obtain the average probability distribution of the degree $j$. The corresponding limiting distribution $\phi_j$ gives the average probability that a vertex is connected by 1, 2, 3 or 4 bond(s) among all the spanning tree configurations. They are rational numbers given as $\phi_1=10957/40464$, $\phi_2=6626035/13636368$, $\phi_3=2943139/13636368$, $\phi_4=124895/4545456$.


\end{abstract}

\maketitle

\section{Introduction}
\label{intro}
The enumeration of the number of spanning trees $N_{ST}(G)$ on a graph $G$ was first considered by Kirchhoff in the analysis of electric circuits \cite{kirchhoff}. It is a problem of fundamental interest in mathematics \cite{bbook,burton93,lyons05,welsh} and physics \cite{temperley72,wu77}. The number of spanning trees corresponds to a special $q \to 0$ limit of the partition function of the $q$-state Potts model in statistical mechanics \cite{fk,wurev}, which in turn is related to the sandpile model \cite{cb,Dhar06}. Just like other limits of the $q$-state Potts model, the spanning tree problem has been investigated intensely for decades, and has various applications in many areas. See, for example, \cite{stbook} and references therein. It is also well known that there is a bijection between close-packed dimer coverings with spanning tree configurations on two related lattices \cite{temperley74}. Some studies on the enumeration of spanning trees and the calculation of their asymptotic growth constants on regular lattices were carried out in Refs. \cite{std,sti,sw,tzengwu}.
Once the total number of spanning trees and its asymptotic growth constant is obtained, the next step is to understand the geometric structure of spanning trees. One interesting question is the probability distribution of the degree of a certain vertex among all the spanning trees \cite{A}. The geometric properties of spanning trees on $\mathbb{Z}^d$ lattices, especially the square lattice, had been considered in \cite{burton93,mdm}. 

Fractals are geometric structures of (generally non-integer) Hausdorff dimension realized by repeated construction of an elementary shape on progressively smaller length scales \cite{Falconer,mandelbrot}. A well-known example of a fractal is the Sierpinski gasket that has been extensively studied in several contexts \cite{Alexander,Daerden,Dhar97,Dhar05,Domany,Gefen80,Gefen81,Gefen8384,Guyer,Kusuoka,Kozak1,Kozak2,Rammal}. Recently, the authors derived rigorously the number of spanning trees on the Sierpinski gasket and conjectured the result for arbitrary dimension \cite{sts}. It is of interest to consider geometric structure of spanning trees on self-similar fractal lattices which have scaling invariance rather than translational invariance.
Different from the lattices that have translational invariance, e.g. the square lattice, the probability distribution of the degree on
Sierpinski gasket depends on the vertex location. Thereby, it is natural to investigate the average of the probability distribution of the degree over all the vertices on $SG(n)$ as $n$ tends to infinite, and compare the values with the corresponding results on the infinite square lattice which is also 4-regular.
In this paper, we shall present such probability distribution of the degree at any given vertex $x$ on the two-dimensional Sierpinski gasket and the average, and the limiting distribution when the number of vertices goes to infinity. 

\setcounter{equation}{0}
\section{Preliminaries}
\label{sectionII}

We first recall some relevant definitions for spanning trees and the Sierpinski gasket in this section. A connected graph (without loops) $G=(V,E)$ is defined by its vertex (site) and edge (bond) sets $V$ and $E$ \cite{bbook,fh}.  Let $v(G)=|V|$ be the number of vertices and $e(G)=|E|$ the number of edges in $G$.  A spanning subgraph $G^\prime$ is a subgraph of $G$ with the same vertex set $V$ and an edge set $E^\prime \subseteq E$. As a tree is a connected graph with no circuits, a spanning tree on $G$ is a spanning subgraph of $G$ that is a tree and hence $e(G') = v(G)-1$. The degree or coordination number $k_i$ of a vertex $v_i \in V$ is the number of edges attached to it.  A $k$-regular graph is a graph with the property that each of its vertices has the same degree $k$. In general, one can associate an edge weight to each edge connecting adjacent vertices $v_i$ and $v_j$ (see, for example \cite{tzengwu}). For simplicity, all edge weights are set to one throughout this paper, so that the weight of each spanning tree is the same. 

The construction of the two-dimensional Sierpinski gasket $SG(n)$ at stage $n$ is shown in Fig. \ref{sgfig}. At stage $n=0$, it is an equilateral triangle; while stage $n+1$ is obtained by the juxtaposition of three $n$-stage structures.  For the two-dimensional Sierpinski gasket $SG(n)$, the numbers of edges and vertices are given by 
\[
e(SG(n)) = 3^{n+1} \ , \qquad 
v(SG(n)) = \frac{3}{2} (3^n+1) \ .
%\label{ev}
\]
Except the three outmost vertices which have degree two, all other vertices of $SG(n)$ have degree four. In the large $n$ limit, $SG$ is $4$-regular. 

\begin{figure}[htbp]
\centering
\unitlength 1.25mm 
%\hspace*{-20mm}
\begin{picture}(96,80)
\put(0,0){\line(1,0){96}}
\multiput(12,20)(48,0){2}{\line(1,0){24}}
\put(24,40){\line(1,0){48}}
\put(36,60){\line(1,0){24}}
\put(0,0){\line(3,5){48}}
\put(48,0){\line(3,5){24}}
\multiput(24,0)(48,0){2}{\line(3,5){12}}
\multiput(24,0)(48,0){2}{\line(-3,5){12}}
\put(48,0){\line(-3,5){24}}
\put(96,0){\line(-3,5){48}}
\put(48,40){\line(3,5){12}}
\put(48,40){\line(-3,5){12}}
\multiput(6,10)(24,0){4}{\line(1,0){12}}
\multiput(18,30)(48,0){2}{\line(1,0){12}}
\multiput(30,50)(24,0){2}{\line(1,0){12}}
\put(42,70){\line(1,0){12}}
\multiput(12,0)(24,0){4}{\line(3,5){6}}
\multiput(24,20)(48,0){2}{\line(3,5){6}}
\multiput(36,40)(24,0){2}{\line(3,5){6}}
\put(48,60){\line(3,5){6}}
\multiput(12,0)(24,0){4}{\line(-3,5){6}}
\multiput(24,20)(48,0){2}{\line(-3,5){6}}
\multiput(36,40)(24,0){2}{\line(-3,5){6}}
\put(48,60){\line(-3,5){6}}
\multiput(3,5)(12,0){8}{\line(1,0){6}}
\multiput(9,15)(24,0){4}{\line(1,0){6}}
\multiput(15,25)(12,0){2}{\line(1,0){6}}
\multiput(63,25)(12,0){2}{\line(1,0){6}}
\multiput(21,35)(48,0){2}{\line(1,0){6}}
\multiput(27,45)(12,0){4}{\line(1,0){6}}
\multiput(33,55)(24,0){2}{\line(1,0){6}}
\multiput(39,65)(12,0){2}{\line(1,0){6}}
\put(45,75){\line(1,0){6}}
\multiput(6,0)(12,0){8}{\line(3,5){3}}
\multiput(12,10)(24,0){4}{\line(3,5){3}}
\multiput(18,20)(12,0){2}{\line(3,5){3}}
\multiput(66,20)(12,0){2}{\line(3,5){3}}
\multiput(24,30)(48,0){2}{\line(3,5){3}}
\multiput(30,40)(12,0){4}{\line(3,5){3}}
\multiput(36,50)(24,0){2}{\line(3,5){3}}
\multiput(42,60)(12,0){2}{\line(3,5){3}}
\put(48,70){\line(3,5){3}}
\multiput(6,0)(12,0){8}{\line(-3,5){3}}
\multiput(12,10)(24,0){4}{\line(-3,5){3}}
\multiput(18,20)(12,0){2}{\line(-3,5){3}}
\multiput(66,20)(12,0){2}{\line(-3,5){3}}
\multiput(24,30)(48,0){2}{\line(-3,5){3}}
\multiput(30,40)(12,0){4}{\line(-3,5){3}}
\multiput(36,50)(24,0){2}{\line(-3,5){3}}
\multiput(42,60)(12,0){2}{\line(-3,5){3}}
\put(48,70){\line(-3,5){3}}
\put(0,-1){\makebox(0,0){$\scriptstyle o$}}
\put(6,-1){\makebox(0,0){$\scriptstyle a_0$}}
\put(12,-1){\makebox(0,0){$\scriptstyle a_1$}}
\put(18,-1){\makebox(0,0){$\scriptstyle a_{11}$}}
\put(24,-1){\makebox(0,0){$\scriptstyle a_2$}}
\put(30,-1){\makebox(0,0){$\scriptstyle a_{210}$}}
\put(36,-1){\makebox(0,0){$\scriptstyle a_{21}$}}
\put(42,-1){\makebox(0,0){$\scriptstyle a_{211}$}}
\put(48,-1){\makebox(0,0){$\scriptstyle a_3$}}
\put(54,-1){\makebox(0,0){$\scriptstyle a_{3100}$}}
\put(60,-1){\makebox(0,0){$\scriptstyle a_{310}$}}
\put(66,-1){\makebox(0,0){$\scriptstyle a_{3101}$}}
\put(72,-1){\makebox(0,0){$\scriptstyle a_{31}$}}
\put(78,-1){\makebox(0,0){$\scriptstyle a_{3110}$}}
\put(84,-1){\makebox(0,0){$\scriptstyle a_{311}$}}
\put(90,-1){\makebox(0,0){$\scriptstyle a_{3111}$}}
\put(96,-1){\makebox(0,0){$\scriptstyle a_4$}}
\put(2,6){\makebox(0,0){$\scriptstyle b_0$}}
\put(10,6){\makebox(0,0){$\scriptstyle c_0$}}
\put(14,6){\makebox(0,0){$\scriptstyle b_{11}$}}
\put(22,6){\makebox(0,0){$\scriptstyle c_{11}$}}
\put(26,7){\makebox(0,0){$\scriptstyle b_{210}$}}
\put(34.5,6){\makebox(0,0){$\scriptstyle c_{210}$}}
\put(38,7){\makebox(0,0){$\scriptstyle b_{211}$}}
\put(46.5,6){\makebox(0,0){$\scriptstyle c_{211}$}}
\put(50,7.5){\makebox(0,0){$\scriptstyle b_{3100}$}}
\put(59,6){\makebox(0,0){$\scriptstyle c_{3100}$}}
\put(62,7.5){\makebox(0,0){$\scriptstyle b_{3101}$}}
\put(71,6){\makebox(0,0){$\scriptstyle c_{3101}$}}
\put(74,7.5){\makebox(0,0){$\scriptstyle b_{3110}$}}
\put(83,6){\makebox(0,0){$\scriptstyle c_{3110}$}}
\put(86,7.5){\makebox(0,0){$\scriptstyle b_{3111}$}}
\put(95,6){\makebox(0,0){$\scriptstyle c_{3111}$}}
\put(5,11){\makebox(0,0){$\scriptstyle b_1$}}
\put(12,8.5){\makebox(0,0){$\scriptstyle \tilde{b}_{11}$}}
\put(19,11){\makebox(0,0){$\scriptstyle c_1$}}
\put(29,11){\makebox(0,0){$\scriptstyle b_{21}$}}
\put(36,9){\makebox(0,0){$\scriptstyle a_{212}$}}
\put(43,11){\makebox(0,0){$\scriptstyle c_{21}$}}
\put(52,11){\makebox(0,0){$\scriptstyle b_{310}$}}
\put(60,9){\makebox(0,0){$\scriptstyle a_{3102}$}}
\put(68,11){\makebox(0,0){$\scriptstyle c_{310}$}}
\put(76,11){\makebox(0,0){$\scriptstyle b_{311}$}}
\put(84,9){\makebox(0,0){$\scriptstyle a_{3112}$}}
\put(92,11){\makebox(0,0){$\scriptstyle c_{311}$}}
\put(8,16){\makebox(0,0){$\scriptstyle \tilde{a}_{11}$}}
\put(16,16){\makebox(0,0){$\scriptstyle \tilde{c}_{11}$}}
\put(31,16){\makebox(0,0){$\scriptstyle b_{212}$}}
\put(41,16){\makebox(0,0){$\scriptstyle c_{212}$}}
\put(55,16){\makebox(0,0){$\scriptstyle b_{3102}$}}
\put(65,16){\makebox(0,0){$\scriptstyle c_{3102}$}}
\put(79,16){\makebox(0,0){$\scriptstyle b_{3112}$}}
\put(89,16){\makebox(0,0){$\scriptstyle c_{3112}$}}
\put(11,21){\makebox(0,0){$\scriptstyle b_2$}}
\put(18,18.5){\makebox(0,0){$\scriptstyle \tilde{b}_{210}$}}
\put(24,18.5){\makebox(0,0){$\scriptstyle \tilde{b}_{21}$}}
\put(30,18.5){\makebox(0,0){$\scriptstyle \tilde{b}_{212}$}}
\put(37,21){\makebox(0,0){$\scriptstyle c_2$}}
\put(59,21){\makebox(0,0){$\scriptstyle b_{31}$}}
\put(66,19){\makebox(0,0){$\scriptstyle a_{3120}$}}
\put(72,19){\makebox(0,0){$\scriptstyle a_{312}$}}
\put(78,19){\makebox(0,0){$\scriptstyle a_{3121}$}}
\put(85,21){\makebox(0,0){$\scriptstyle c_{31}$}}
\put(13,26){\makebox(0,0){$\scriptstyle \tilde{a}_{210}$}}
\put(22.5,26){\makebox(0,0){$\scriptstyle \tilde{c}_{210}$}}
\put(26,27){\makebox(0,0){$\scriptstyle \tilde{a}_{212}$}}
\put(35,26){\makebox(0,0){$\scriptstyle \tilde{c}_{212}$}}
\put(61,26){\makebox(0,0){$\scriptstyle b_{3120}$}}
\put(71,26){\makebox(0,0){$\scriptstyle c_{3120}$}}
\put(74,27.5){\makebox(0,0){$\scriptstyle b_{3121}$}}
\put(83,26){\makebox(0,0){$\scriptstyle c_{3121}$}}
\put(17,31){\makebox(0,0){$\scriptstyle \tilde{a}_{21}$}}
\put(24,29){\makebox(0,0){$\scriptstyle \tilde{b}_{211}$}}
\put(31,31){\makebox(0,0){$\scriptstyle \tilde{c}_{21}$}}
\put(64,31){\makebox(0,0){$\scriptstyle b_{312}$}}
\put(72,29){\makebox(0,0){$\scriptstyle a_{3122}$}}
\put(80,31){\makebox(0,0){$\scriptstyle c_{312}$}}
\put(19,36){\makebox(0,0){$\scriptstyle \tilde{a}_{211}$}}
\put(29,36){\makebox(0,0){$\scriptstyle \tilde{c}_{211}$}}
\put(67,36){\makebox(0,0){$\scriptstyle b_{3122}$}}
\put(77,36){\makebox(0,0){$\scriptstyle c_{3122}$}}
\put(23,41){\makebox(0,0){$\scriptstyle b_3$}}
\put(73,41){\makebox(0,0){$\scriptstyle c_3$}}
\put(47,81){\makebox(0,0){$\scriptstyle b_4$}}
\end{picture}
\vspace*{3mm}
\caption{\footnotesize{The notation for the vertices of the Sierpinski gasket $SG(4)$. The vertices $\tilde{x}_{\vec{\gamma}}$ inside the triangle $(b_3,c_3,b_4)$ are reflection of the vertices $x_{\vec{\gamma}}$ inside the triangle $(a_3,c_3,a_4)$ with respect to the line connecting $o$ and $c_3$, and are not shown.}} 
\label{sgvertex}
\end{figure}
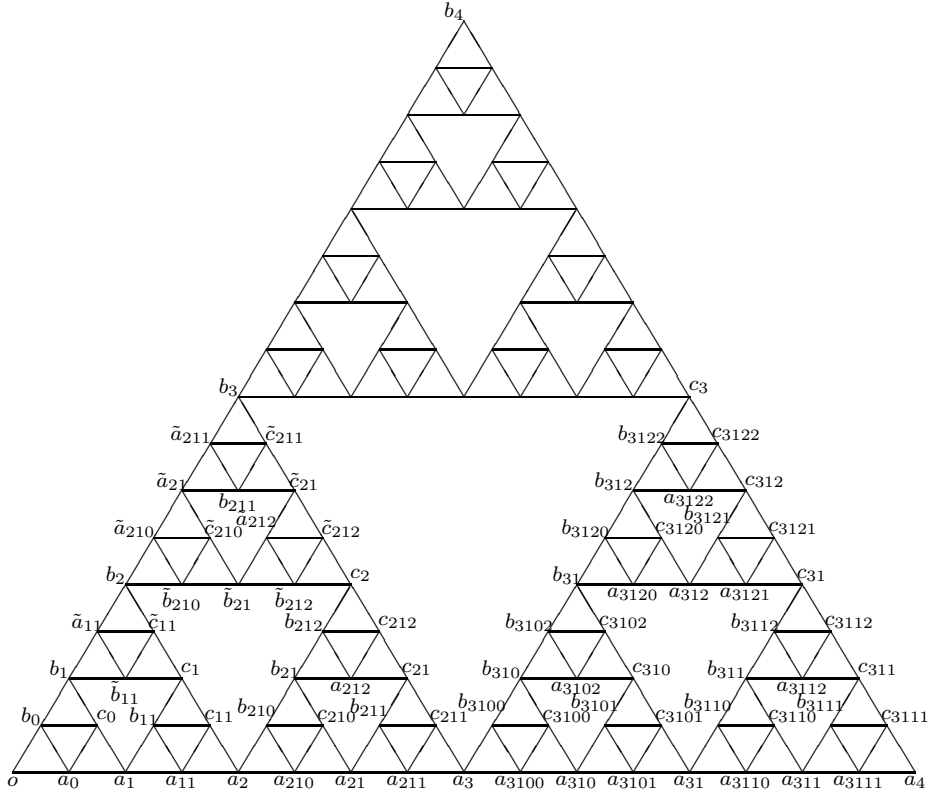

Let us define the notation for the vertices of $SG(n)$ to be used. An illustration for $SG(4)$ is shown in Fig. \ref{sgvertex}. The denotation of the vertices is given progressively with increasing number of digits in the subscript as follows. First of all, fix $o$ as the leftmost vertex. Consider the $SG(m)$ with $0 \le m \le n$ always has $o$ as its leftmost vertex, and denote $a_m$ and $b_m$ as its rightmost and topmost vertices, respectively. $c_m$ is defined such that the vertices $a_m$, $b_m$ and $c_m$ demarcate the largest lacunary triangle of $SG(m+1)$. We then define the vertex in the middle of the line connecting $a_m$ and $a_{m+1}$ with $m \ge 1$ as $a_{m,1}$. Similarly, $a_{m,1}$ and the associated $b_{m,1}$ and $c_{m,1}$ demarcate a lacunary triangle with $b_{m,1}$ on the left and $c_{m,1}$ on the right. Next for $m \ge 2$, we append the subscript $m,1,0$ for the vertices of the largest lacunary inside the triangle with outmost vertices $a_m$, $a_{m,1}$, $b_{m,1}$; the subscript $m,1,1$ for the vertices of the largest lacunary inside the triangle with outmost vertices $a_{m,1}$, $a_{m+1}$, $c_{m,1}$; the subscript $m,1,2$ for the vertices of the largest lacunary inside the triangle with outmost vertices $b_{m,1}$, $c_{m,1}$, $c_m$, etc. In general for the vertices of $SG(n)$, we use the notation $x_{\vec{\gamma}}$ where $x=a,b,c$ and the subscript $\vec{\gamma} = (\gamma_1,...,\gamma_s)$ has $s$ components with $1 \leq s \leq n$, $1 \leq \gamma_1 < n$ and $\gamma_k \in \{0,1,2\}$ for $k \in \{2,3,...,s\}$.  
For the vertices above the extended line connecting $o$ and $c_0$, we will also use the notation $\tilde{x}_{\vec{\gamma}}$ such that it is the reflection of the vertex $x_{\vec{\gamma}}$ with respect to this line. For examples, $a_{22}=\tilde{b}_{21}$, $b_{221}=\tilde{a}_{212}$, $c_{222}=\tilde{c}_{211}$, etc. The advantage of such vertex notation is that the quantities to be studied for the vertices $x_{\gamma_1,...,\gamma_s}$ with $s \ge 2$ components in the subscript can be expressed in terms of the quantities for the vertices with $s-1$ components in the subscript as shown in Section \ref{sectionV}.

\begin{figure}[htbp]
\centering
\unitlength 0.9mm \hspace*{3mm}
\begin{picture}(108,40)
\put(0,0){\line(1,0){6}}
\put(0,0){\line(3,5){3}}
\put(6,0){\line(-3,5){3}}
\put(3,-4){\makebox(0,0){$SG(0)$}}
\put(12,0){\line(1,0){12}}
\put(12,0){\line(3,5){6}}
\put(24,0){\line(-3,5){6}}
\put(15,5){\line(1,0){6}}
\put(18,0){\line(3,5){3}}
\put(18,0){\line(-3,5){3}}
\put(18,-4){\makebox(0,0){$SG(1)$}}
\put(30,0){\line(1,0){24}}
\put(30,0){\line(3,5){12}}
\put(54,0){\line(-3,5){12}}
\put(36,10){\line(1,0){12}}
\put(42,0){\line(3,5){6}}
\put(42,0){\line(-3,5){6}}
\multiput(33,5)(12,0){2}{\line(1,0){6}}
\multiput(36,0)(12,0){2}{\line(3,5){3}}
\multiput(36,0)(12,0){2}{\line(-3,5){3}}
\put(39,15){\line(1,0){6}}
\put(42,10){\line(3,5){3}}
\put(42,10){\line(-3,5){3}}
\put(42,-4){\makebox(0,0){$SG(2)$}}
\put(60,0){\line(1,0){48}}
\put(72,20){\line(1,0){24}}
\put(60,0){\line(3,5){24}}
\put(84,0){\line(3,5){12}}
\put(84,0){\line(-3,5){12}}
\put(108,0){\line(-3,5){24}}
\put(66,10){\line(1,0){12}}
\put(90,10){\line(1,0){12}}
\put(78,30){\line(1,0){12}}
\put(72,0){\line(3,5){6}}
\put(96,0){\line(3,5){6}}
\put(84,20){\line(3,5){6}}
\put(72,0){\line(-3,5){6}}
\put(96,0){\line(-3,5){6}}
\put(84,20){\line(-3,5){6}}
\multiput(63,5)(12,0){4}{\line(1,0){6}}
\multiput(66,0)(12,0){4}{\line(3,5){3}}
\multiput(66,0)(12,0){4}{\line(-3,5){3}}
\multiput(69,15)(24,0){2}{\line(1,0){6}}
\multiput(72,10)(24,0){2}{\line(3,5){3}}
\multiput(72,10)(24,0){2}{\line(-3,5){3}}
\multiput(75,25)(12,0){2}{\line(1,0){6}}
\multiput(78,20)(12,0){2}{\line(3,5){3}}
\multiput(78,20)(12,0){2}{\line(-3,5){3}}
\put(81,35){\line(1,0){6}}
\put(84,30){\line(3,5){3}}
\put(84,30){\line(-3,5){3}}
\put(84,-4){\makebox(0,0){$SG(3)$}}
\end{picture}

\vspace*{5mm}
\caption{\footnotesize{The first four stages $n=0,1,2,3$ of the two-dimensional Sierpinski gasket $SG(n)$.}} 
\label{sgfig}
\end{figure}
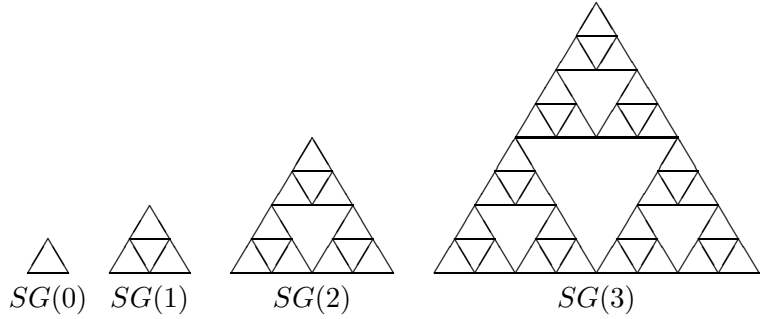

Let us define the following quantities as in \cite{sts}.

\begin{defi} 
\label{defisg2}
Consider the two-dimensional Sierpinski gasket $SG(n)$ at stage $n$. (i) Define $f(n) \equiv N_{ST}(SG(n))$ as the number of spanning trees. (ii) Define $g(n)$ as the number of spanning subgraphs with two trees such that the vertex $b_n$ belongs to one tree and the set of vertices $\{o, a_n\}$ belong to the other tree. (iii) Define $h(n)$ as the number of spanning subgraphs with three trees such that each of the outmost vertices $o$, $a_n$ and $b_n$ belongs to a different tree. 
\end{defi}

Notice that for the spanning subgraph configurations counted by $g(n)$, it is possible that the vertex $b_n$ is an isolated vertex with no bonds of trees connecting to it. A similar statement applies to the outmost verticex $o$, $a_n$, $b_n$ for the spanning subgraph configurations counted by $h(n)$. For a given vertex, we would like to investigate the number of bonds of spanning trees connecting to it among all the spanning tree configurations. We have the following definitions.

\begin{defi} 
\label{defisg2n}
Consider the two-dimensional Sierpinski gasket $SG(n)$ at stage $n$. For a certain vertex $x \in V(SG(n))$, the number of bond(s) connects to it in a spanning tree configuration is denoted as $j \in \{1,2,3,4\}$ or $i \in \{0,1,2,3,4\}$. (i) Define $f_j(n,x)$ as the number of spanning trees such that there is (are) $j$ bond(s) connecting the vertex $x$. Define the probability $F_j(n,x) = f_j(n,x)/f(n)$. (ii) Define $g_i(n,x)$ as the number of spanning subgraphs with two trees such that the vertex $b_n$ belongs to one tree and the set of vertices $\{o, a_n\}$ belongs to the other tree, and there is (are) $i$ bond(s) connecting the vertex $x$. Define the probability $G_i(n,x)=g_i(n,x)/g(n)$. (iii) Define $h_i(n,x)$ as the number of spanning subgraphs with three trees such that each of the outmost vertices $o$, $a_n$, $b_n$ belongs to a different tree, and there is (are) $i$ bond(s) connecting the vertex $x$. Define the probability $H_i(n,x)=h_i(n,x)/h(n)$.
\end{defi}

For any vertex $x$ of $SG(n)$, the following relations for the probabilities should be satisfied,
\[
\sum_{j=1}^4 F_j(n,x) = \sum_{i=0}^4 G_j(n,x) = \sum_{i=0}^4 H_j(n,x) = 1 \ , 
\]
which serves as a check for the results obtained.

In this paper, we derive rigorously $F_j(n,x)$ or $f_j(n,x)$ for an arbitrary vertex $x \in V(SG(n))$ with $j=1,2,3,4$. 
Such probability on translational invariance lattices in the infinite-vertex limit is independent of the vertex location. In contrast, as the Sierpinski gasket is a self-similar fractal lattice which has scaling invariance rather than translational invariance, our results depend on the location of $x$. 
We shall consider the simplest vertex $x=o$ to obtain $F_j(n,o)$ as Theorem \ref{theoremFG} and its infinite $n$ limit as Corollary \ref{corollaryf} in Section \ref{sectionIII}, then move on to the vertices $x \in \{a_m, b_m, c_m\}$ with $0 \le m < n$ to have Theorem \ref{theoremF} and Corollary \ref{corollaryac} in Section \ref{sectionIV}. $F_j(n,x)$ for the rest vertices will be treated in Section \ref{sectionV} as Propositions \ref{propositionF} and \ref{propositionFF}. The summation and average of all $F_j(n,x)$ for a given stage $n$ will be studied in Section \ref{sectionVI}, and such average in the infinite $n$ limit will be obtained as Theorem \ref{theoremphi}.
 
\setcounter{equation}{0}
\section{$F_j(n,o)$ with $j \in \{1,2\}$}
\label{sectionIII}
Consider the Sierpinski gasket $SG(n)$ at stage $n$. We will derive $F_j(n,x)$ for the vertex $x=o$ in this section. Since the leftmost vertex $o$ has degree two, $f_j(n,o)=0$ for $j=3,4$ and any $n \ge 0$. Similarly, we only need $i \in \{0,1,2\}$ for $g_i(n,x)$ and 
$h_i(n,x)$ with $x \in \{o, a_n, b_n\}$. Due to the symmetry of $SG(n)$, we have $f_j(n,o) = f_j(n,a_n) = f_j(n,b_n)$, $g_j(n,o) = g_j(n,a_n)$ with $j=1,2$, and $h_i(n,o) = h_i(n,a_n) = h_i(n,b_n)$ with $i=0,1,2$. According to the definition, $g_0(n,o) = g_0(n,a_n) = 0$, but $g_0(n,b_n) \neq 0$ for any $n \geq 0$. In fact, $g_0(n,b_n)$ is the only $g_0(n,x)$ with non-zero value, and $h_0(n,x)$ is non-zero only when $x \in \{o,a_n,b_n\}$. The initial values for $x=o$ at stage $n=0$ are $f(0)=3$ with decompositions $f_1(0,o)=2$ and $f_2(0,o)=1$, $g(0)=1$ with decompositions $g_1(0,o)=1$ and $g_2(0,o)=0$, $h(0)=1$ with decompositions $h_0(0,o)=1$ and $h_j(0,o)=0$ for $j=1,2$. We also have $g_0(0,b_0)=1$ and $g_j(0,b_0)=0$ for $j=1,2$.

\begin{figure}[htbp]
\unitlength 1mm 
\begin{picture}(94,12)
\multiput(0,0)(1,0){13}{\circle*{0.2}}
\multiput(0,0)(0.6,1){11}{\circle*{0.2}}
\multiput(12,0)(-0.6,1){11}{\circle*{0.2}}
\put(15,5){\makebox(0,0){$=$}}
\put(18,0){\line(1,0){6}}
\put(18,0){\line(3,5){3}}
\multiput(24,0)(6,0){2}{\line(-3,5){3}}
\multiput(24,0)(1,0){7}{\circle*{0.2}}
\multiput(21,5)(1,0){7}{\circle*{0.2}}
\multiput(24,0)(0.6,1){6}{\circle*{0.2}}
\multiput(21,5)(0.6,1){6}{\circle*{0.2}}
\multiput(27,5)(-0.6,1){6}{\circle*{0.2}}
\put(31,5){\makebox(0,0){$\times 3$}}
\put(36,5){\makebox(0,0){$+$}}
\put(39,0){\line(1,0){6}}
\put(45,0){\line(-3,5){3}}
\multiput(39,0)(6,0){2}{\line(3,5){3}}
\multiput(45,0)(1,0){7}{\circle*{0.2}}
\multiput(42,5)(1,0){7}{\circle*{0.2}}
\multiput(42,5)(0.6,1){6}{\circle*{0.2}}
\multiput(51,0)(-0.6,1){11}{\circle*{0.2}}
\put(52,5){\makebox(0,0){$\times 3$}}
\put(57,5){\makebox(0,0){$+$}}
\put(60,0){\line(1,0){6}}
\put(60,0){\line(3,5){3}}
\multiput(66,0)(3,5){2}{\line(-3,5){3}}
\multiput(66,0)(1,0){7}{\circle*{0.2}}
\multiput(63,5)(1,0){7}{\circle*{0.2}}
\multiput(66,0)(0.6,1){6}{\circle*{0.2}}
\multiput(63,5)(0.6,1){6}{\circle*{0.2}}
\multiput(72,0)(-0.6,1){6}{\circle*{0.2}}
\put(73,5){\makebox(0,0){$\times 3$}}
\put(78,5){\makebox(0,0){$+$}}
\put(81,0){\line(3,5){3}}
\put(87,0){\line(-3,5){3}}
\multiput(81,0)(3,5){2}{\line(1,0){6}}
\multiput(87,0)(1,0){7}{\circle*{0.2}}
\multiput(87,0)(0.6,1){6}{\circle*{0.2}}
\multiput(84,5)(0.6,1){6}{\circle*{0.2}}
\multiput(93,0)(-0.6,1){11}{\circle*{0.2}}
\put(94,5){\makebox(0,0){$\times 3$}}
\end{picture}

\begin{picture}(94,12)
\put(15,5){\makebox(0,0){$+$}}
\put(24,0){\line(3,5){3}}
\multiput(24,0)(3,5){2}{\line(-3,5){3}}
\multiput(18,0)(1,0){13}{\circle*{0.2}}
\multiput(18,0)(0.6,1){11}{\circle*{0.2}}
\multiput(21,5)(1,0){7}{\circle*{0.2}}
\multiput(30,0)(-0.6,1){6}{\circle*{0.2}}
\put(31,5){\makebox(0,0){$\times 3$}}
\put(36,5){\makebox(0,0){$+$}}
\put(45,0){\line(-3,5){3}}
\multiput(45,0)(-3,5){2}{\line(3,5){3}}
\multiput(39,0)(1,0){13}{\circle*{0.2}}
\multiput(51,0)(-0.6,1){11}{\circle*{0.2}}
\multiput(45,0)(1,0){7}{\circle*{0.2}}
\multiput(42,5)(1,0){7}{\circle*{0.2}}
\multiput(39,0)(0.6,1){6}{\circle*{0.2}}
\put(52,5){\makebox(0,0){$\times 3$}}
\put(57,5){\makebox(0,0){$+$}}
\put(63,5){\line(3,5){3}}
\multiput(66,0)(6,0){2}{\line(-3,5){3}}
\multiput(60,0)(1,0){13}{\circle*{0.2}}
\multiput(63,5)(1,0){7}{\circle*{0.2}}
\multiput(60,0)(0.6,1){6}{\circle*{0.2}}
\multiput(66,0)(0.6,1){6}{\circle*{0.2}}
\multiput(69,5)(-0.6,1){6}{\circle*{0.2}}
\put(73,5){\makebox(0,0){$\times 3$}}
\put(78,5){\makebox(0,0){$+$}}
\put(90,5){\line(-3,5){3}}
\multiput(81,0)(6,0){2}{\line(3,5){3}}
\multiput(81,0)(1,0){13}{\circle*{0.2}}
\multiput(84,5)(1,0){7}{\circle*{0.2}}
\multiput(84,5)(0.6,1){6}{\circle*{0.2}}
\multiput(87,0)(-0.6,1){6}{\circle*{0.2}}
\multiput(93,0)(-0.6,1){6}{\circle*{0.2}}
\put(94,5){\makebox(0,0){$\times 3$}}
\end{picture}

\begin{picture}(94,12)
\put(15,5){\makebox(0,0){$+$}}
\put(18,0){\line(1,0){6}}
\put(21,5){\line(3,5){3}}
\put(30,0){\line(-3,5){3}}
\multiput(18,0)(0.6,1){6}{\circle*{0.2}}
\multiput(24,0)(0.6,1){6}{\circle*{0.2}}
\multiput(24,0)(1,0){7}{\circle*{0.2}}
\multiput(21,5)(1,0){7}{\circle*{0.2}}
\multiput(24,0)(-0.6,1){6}{\circle*{0.2}}
\multiput(27,5)(-0.6,1){6}{\circle*{0.2}}
\put(36,5){\makebox(0,0){$+$}}
\put(45,0){\line(1,0){6}}
\put(39,0){\line(3,5){3}}
\put(48,5){\line(-3,5){3}}
\multiput(39,0)(1,0){7}{\circle*{0.2}}
\multiput(42,5)(1,0){7}{\circle*{0.2}}
\multiput(42,5)(0.6,1){6}{\circle*{0.2}}
\multiput(45,0)(0.6,1){6}{\circle*{0.2}}
\multiput(45,0)(-0.6,1){6}{\circle*{0.2}}
\multiput(51,0)(-0.6,1){6}{\circle*{0.2}}
\end{picture}

\caption{\footnotesize{Illustration for the expression of  $h(n+1)$. The multiplication for the eight configurations on the right-hand-side corresponds to three possible orientations.}} 
\label{hfig}
\end{figure}
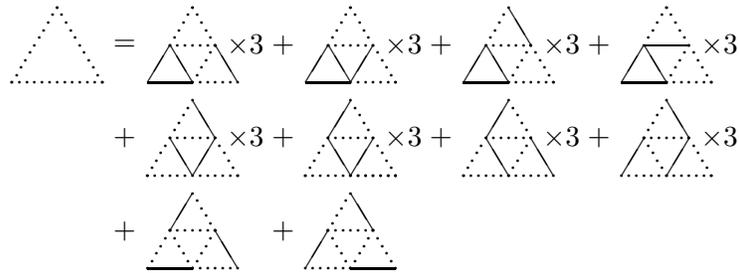

The following recursion relations was derived in \cite{sts} for $n \ge 0$,
\beqs
\left\{\begin{array}{lll}
f(n+1) & = & 6f(n)^2g(n) \ , \cr
g(n+1) & = & f(n)^2h(n) + 7f(n)g(n)^2 \ , \cr
h(n+1) & = & 12f(n)g(n)h(n) + 14g(n)^3 \ ,
\end{array}
\right.
\label{fgh}
\eeqs
as illustrated in Figs. \ref{ffig}-\ref{hfig}. $f(n)$, $g(n)$, $h(n)$ were solved exactly in \cite{sts} such that they satisfy the relation $3g(n)^2=f(n)h(n)$. It follows that the second and third lines of (\ref{fgh}) can be simplified as
\beqs
\left\{\begin{array}{lll}
g(n+1) & = & 10f(n)g(n)^2 = \frac{10}{3}f(n)^2h(n) \ , \cr
h(n+1) & = & 50g(n)^3 = \frac{50}{3}f(n)g(n)h(n) \ .
\end{array}
\right.
\label{gh}
\eeqs

\begin{figure}[htbp]
\unitlength 1mm 
\begin{picture}(120,12)
\put(0,0){\line(1,0){12}}
\put(0,0){\line(3,5){6}}
\put(12,0){\line(-3,5){6}}
\put(15,5){\makebox(0,0){$=$}}
\put(18,0){\line(1,0){6}}
\put(21,5){\line(1,0){6}}
\put(18,0){\line(3,5){6}}
\put(30,0){\line(-3,5){6}}
\put(24,0){\line(-3,5){3}}
\multiput(24,0)(1,0){7}{\circle*{0.2}}
\multiput(24,0)(0.6,1){6}{\circle*{0.2}}
\put(33,5){\makebox(0,0){$+$}}
\put(36,0){\line(1,0){12}}
\put(39,5){\line(1,0){6}}
\put(36,0){\line(3,5){6}}
\multiput(42,0)(3,5){2}{\line(-3,5){3}}
\multiput(42,0)(0.6,1){6}{\circle*{0.2}}
\multiput(48,0)(-0.6,1){6}{\circle*{0.2}}
\put(51,5){\makebox(0,0){$+$}}
\put(54,0){\line(1,0){12}}
\put(54,0){\line(3,5){6}}
\put(60,0){\line(3,5){3}}
\multiput(60,0)(6,0){2}{\line(-3,5){3}}
\multiput(57,5)(1,0){7}{\circle*{0.2}}
\multiput(63,5)(-0.6,1){6}{\circle*{0.2}}
\put(69,5){\makebox(0,0){$+$}}
\put(72,0){\line(1,0){12}}
\put(84,0){\line(-3,5){6}}
\put(78,0){\line(-3,5){3}}
\multiput(72,0)(6,0){2}{\line(3,5){3}}
\multiput(75,5)(1,0){7}{\circle*{0.2}}
\multiput(75,5)(0.6,1){6}{\circle*{0.2}}
\put(87,5){\makebox(0,0){$+$}}
\put(90,0){\line(1,0){12}}
\put(102,0){\line(-3,5){6}}
\put(93,5){\line(1,0){6}}
\multiput(96,0)(-3,5){2}{\line(3,5){3}}
\multiput(90,0)(0.6,1){6}{\circle*{0.2}}
\multiput(96,0)(-0.6,1){6}{\circle*{0.2}}
\put(105,5){\makebox(0,0){$+$}}
\put(120,0){\line(-3,5){6}}
\put(108,0){\line(3,5){6}}
\put(114,0){\line(3,5){3}}
\multiput(114,0)(-3,5){2}{\line(1,0){6}}
\multiput(108,0)(1,0){7}{\circle*{0.2}}
\multiput(114,0)(-0.6,1){6}{\circle*{0.2}}
\end{picture}

\caption{\footnotesize{Illustration for the expression of  $f(n+1)$.}} 
\label{ffig}
\end{figure}
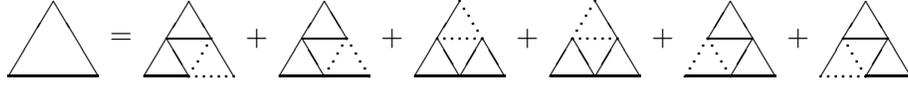

\begin{figure}[htbp]
\unitlength 1mm 
\begin{picture}(84,12)
\put(0,0){\line(1,0){12}}
\multiput(0,0)(0.6,1){11}{\circle*{0.2}}
\multiput(12,0)(-0.6,1){11}{\circle*{0.2}}
\put(15,5){\makebox(0,0){$=$}}
\put(18,0){\line(1,0){12}}
\multiput(18,0)(6,0){2}{\line(3,5){3}}
\multiput(24,0)(6,0){2}{\line(-3,5){3}}
\multiput(21,5)(1,0){7}{\circle*{0.2}}
\multiput(21,5)(0.6,1){6}{\circle*{0.2}}
\multiput(27,5)(-0.6,1){6}{\circle*{0.2}}
\put(33,5){\makebox(0,0){$+$}}
\put(36,0){\line(1,0){12}}
\put(39,5){\line(1,0){6}}
\put(39,5){\line(3,5){3}}
\put(45,5){\line(-3,5){3}}
\multiput(36,0)(0.6,1){6}{\circle*{0.2}}
\multiput(42,0)(0.6,1){6}{\circle*{0.2}}
\multiput(42,0)(-0.6,1){6}{\circle*{0.2}}
\multiput(48,0)(-0.6,1){6}{\circle*{0.2}}
\put(51,5){\makebox(0,0){$+$}}
\put(54,0){\line(1,0){12}}
\put(57,5){\line(1,0){6}}
\put(54,0){\line(3,5){3}}
\put(60,0){\line(-3,5){3}}
\multiput(57,5)(0.6,1){6}{\circle*{0.2}}
\multiput(60,0)(0.6,1){6}{\circle*{0.2}}
\multiput(66,0)(-0.6,1){11}{\circle*{0.2}}
\put(69,5){\makebox(0,0){$+$}}
\put(72,0){\line(1,0){12}}
\put(75,5){\line(1,0){6}}
\put(78,0){\line(3,5){3}}
\put(84,0){\line(-3,5){3}}
\multiput(72,0)(0.6,1){11}{\circle*{0.2}}
\multiput(78,0)(-0.6,1){6}{\circle*{0.2}}
\multiput(81,5)(-0.6,1){6}{\circle*{0.2}}
\end{picture}

\begin{picture}(84,12)
\put(15,5){\makebox(0,0){$+$}}
\put(18,0){\line(3,5){3}}
\multiput(18,0)(3,5){2}{\line(1,0){6}}
\multiput(24,0)(6,0){2}{\line(-3,5){3}}
\multiput(24,0)(1,0){7}{\circle*{0.2}}
\multiput(24,0)(0.6,1){6}{\circle*{0.2}}
\multiput(21,5)(0.6,1){6}{\circle*{0.2}}
\multiput(27,5)(-0.6,1){6}{\circle*{0.2}}
\put(33,5){\makebox(0,0){$+$}}
\put(48,0){\line(-3,5){3}}
\multiput(36,0)(6,0){2}{\line(3,5){3}}
\multiput(39,5)(3,-5){2}{\line(1,0){6}}
\multiput(36,0)(1,0){7}{\circle*{0.2}}
\multiput(42,0)(-0.6,1){6}{\circle*{0.2}}
\multiput(39,5)(0.6,1){6}{\circle*{0.2}}
\multiput(45,5)(-0.6,1){6}{\circle*{0.2}}
\put(51,5){\makebox(0,0){$+$}}
\put(54,0){\line(1,0){12}}
\put(54,0){\line(3,5){3}}
\multiput(60,0)(3,5){2}{\line(-3,5){3}}
\multiput(57,5)(1,0){7}{\circle*{0.2}}
\multiput(57,5)(0.6,1){6}{\circle*{0.2}}
\multiput(60,0)(0.6,1){6}{\circle*{0.2}}
\multiput(66,0)(-0.6,1){6}{\circle*{0.2}}
\put(69,5){\makebox(0,0){$+$}}
\put(72,0){\line(1,0){12}}
\put(84,0){\line(-3,5){3}}
\multiput(78,0)(-3,5){2}{\line(3,5){3}}
\multiput(75,5)(1,0){7}{\circle*{0.2}}
\multiput(72,0)(0.6,1){6}{\circle*{0.2}}
\multiput(78,0)(-0.6,1){6}{\circle*{0.2}}
\multiput(81,5)(-0.6,1){6}{\circle*{0.2}}
\end{picture}

\caption{\footnotesize{Illustration for the expression of  $g(n+1)$.}} 
\label{gfig}
\end{figure}
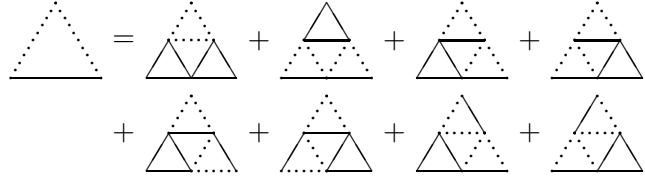

Using Figs. \ref{ffig}-\ref{hfig} for vertex $o$, we obtain the following recursion relations for $j=1,2$:

\beqs
\left\{ \begin{array}{lll}
f_j(n+1,o) & = & 4f_j(n,o) f(n) g(n) + 2g_{j}(n,o)f(n)^2 \ , \cr
g_{j}(n+1,o) & = & f_j(n,o)f(n)h(n) + 3f_j(n,o)g(n)^2 + 4g_{j}(n,o)f(n)g(n) \ ,
\end{array}
\right.
\label{fgj}
\eeqs
and
\beqs
\left\{ \begin{array}{lll}
g_{j}(n+1,b_{n+1}) & = & h_j(n,o)f(n)^2 + f_{j}(n,o)g(n)^2 + 2[g_{j}(n,o) + 2g_{j}(n,b_n)]f(n)g(n) \ , \cr
h_j(n+1,o) & = & 4h_j(n,o) f(n) g(n) + 6g_{j}(n,b_n)g(n)^2
+ 4f_j(n,o)g(n)h(n) \\ 
& & + 8g_{j}(n,o)g(n)^2 + 2[g_{j}(n,o) + g_{j}(n,b_n)]f(n)h(n) \ .
\end{array}
\right.
\label{ghj}
\eeqs
Setting $f_0(n,o)=0$ and $g_0(n,o)=0$, (\ref{ghj}) reduces to
\beqs
\left\{ \begin{array}{l}
g_{0}(n+1,b_{n+1}) = h_0(n,o)f(n)^2 + 4g_{0}(n,b_n)f(n)g(n) \ , \cr
h_0(n+1,o) = 4h_0(n,o)f(n)g(n) + 6g_{0}(n,b_n)g(n)^2 + 2g_{0}(n,b_n)f(n)h(n) \ . 
\end{array} \right.
\label{gh0}
\eeqs
The initial values for the probabilities are 
\beqs
\left\{ \begin{array}{l}
F_1(0,o)=2/3 \ , \ F_2(0,o)=1/3 \ , \ G_{1}(0,o)=1 \ , \ G_{2}(0,o)=0 , \cr
H_0(0,o)=1 \ , \ G_{0}(0,b_0)=1 \ , \ G_{j}(0,b_0)=H_j(0,o)=0 \ \mbox{with} \ j=1,2 \ .
\end{array} \right.
\label{iv0}
\eeqs
Divide the quantities in (\ref{fgj})-(\ref{gh0}) by $f(n+1)$, $g(n+1)$ or $h(n+1)$ given in (\ref{fgh}) or (\ref{gh}), we get
\beqs
\left\{ \begin{array}{lll}
F_j(n+1,o) & = & \frac 23 F_j(n,o) + \frac 13 G_{j}(n,o) \ , \cr
G_{j}(n+1,o) & = & \frac 35 F_j(n,o) + \frac 25 G_{j}(n,o) \ ,
\end{array}
\right.
\label{FGj}
\eeqs
\beqs
\left\{ \begin{array}{lll}
G_{j}(n+1,b_{n+1}) & = & \frac{2}{5}G_{j}(n,b_n) + \frac{3}{10}H_j(n,o) + \frac {1}{10}F_{j}(n,o) + \frac {1}{5}G_{j}(n,o) \ , \cr
H_j(n+1,o) & = &  \frac{6}{25}G_{j}(n,b_n) + \frac{6}{25}H_j(n,o) + \frac{6}{25}F_j(n,o) + \frac{7}{25}G_{j}(n,o) \ ,
\end{array}
\right.
\label{GHj}
\eeqs
for $j=1,2$, and
\beqs
\left\{ \begin{array}{lll}
G_{0}(n+1,b_{n+1})  & = & \frac25 G_{0}(n,b_n)+\frac3{10} H_0(n,o) \ , \cr
H_0(n+1,o) & = & \frac{6}{25}G_{0}(n,b_n) + \frac{6}{25}H_0(n,o) \ .
\end{array} \right.
\label{GH0}
\eeqs
The probabilities $F_j(n,o)$ and $G_j(n,o)$ with $j=1,2$ can be solved exactly by linear algebra as follows.

\begin{theo}
\label{theoremFG} 
For the Sierpinski gasket $SG(n)$ with non-negative integer $n$,
\beqn
F_1(n,o) & = & \frac{11}{14} - \frac5{42}\bigl(\frac 1{15}\bigr)^n \ , \quad G_{1}(n,o) = \frac{11}{14} + \frac {3}{14}\bigl(\frac 1{15}\bigr)^n \ , \cr\cr
F_2(n,o) & = & \frac{3}{14} + \frac5{42}\bigl(\frac 1{15}\bigr)^n \ , \quad G_{2}(n,o) = \frac {3}{14} - \frac {3}{14}\bigl(\frac 1{15}\bigr)^n \ .
\eeqn
\end{theo}

{\sl Proof} \quad  
Denote the vector $V_j(n)=(F_j(n,o),G_{j}(n,o))^T$ with $j=1,2$ and $n \ge 0$. By (\ref{iv0}) and (\ref{FGj}), we have
\[
V_j(n) = A^n V_j(0) \ ,
\]
where the matrix $A$ and the initial $V_j(0)$ are
\[
A=\left[\begin{array}{cc}
\frac 23 & \frac 13 \cr
\frac 35 & \frac 25
\end{array}
\right] \quad \mbox{and} \quad 
V_1(0)=\left[\begin{array}{cc}
\frac 23 \cr
1
\end{array}
\right] \ , \quad 
V_2(0)=\left[\begin{array}{cc}
\frac 13 \cr
0
\end{array}
\right] \ .
\]
The matrix $A$ can be diagonalized such that
\[
V_j(n) = Q_A D_A^n Q^{-1}_A V_j(0) \ ,
\]
where
\beqn
D_A = \left[\begin{array}{cc}
1 & 0\\
0 & \frac 1{15}
\end{array}
\right] 
\quad \mbox{and} \quad 
Q_A=\left[\begin{array}{cc}
1 & 5\\
1 & -9
\end{array}
\right] \ .
\eeqn
Therefore,
\[
V_j(n) = Q_A D_A^n Q_A^{-1} V_j(0) = \left[ \begin{array}{cc}
\frac{9}{14} + \frac {5}{14} \bigl(\frac {1}{15}\bigr)^n & \frac{5}{14} - \frac {5}{14}\bigl(\frac {1}{15}\bigr)^n \cr
\frac{9}{14} - \frac {9}{14} \bigl(\frac {1}{15}\bigr)^n & \frac5{14} + \frac 9{14}\bigl(\frac {1}{15}\bigr)^n
\end{array} \right] V_j(0)
\]
for $j=1,2$, and the proof is completed. 
\ $\Box$

From Theorem \ref{theoremFG} and the exact expressions of $f(n)$, $g(n)$, $h(n)$ in \cite{sts}, we have the following corollary.

\begin{cor}
\label{corollaryf} 
For the Sierpinski gasket $SG(n)$ with non-negative integer $n$,
\beqn
f_{1}(n,o) & = & \Bigl[\frac{11}{14} - \frac{5}{42} \bigl(\frac1{15}\bigr)^n\Bigr] \Bigl[2^{\alpha(n)}3^{\beta(n)}5^{\gamma(n)}\Bigr] \ , \cr\cr
f_2(n,o) & = & \Bigl[\frac{3}{14} + \frac5{42} \bigl(\frac1{15}\bigr)^n\Bigr] \Bigl[2^{\alpha(n)}3^{\beta(n)}5^{\gamma(n)}\Bigr] \ ,
\eeqn
where $\alpha(n)=\frac 12(3^n-1)$, $\beta(n)=\frac 14(3^{n+1}+2n+1)$ and $\gamma(n)=\frac 14(3^n-2n-1)$.
The limiting probabilities for the vertex $o$ are
\beqn
\lim_{n\rightarrow\infty}F_1(n,o) = \frac{11}{14} \ , \qquad \lim_{n\rightarrow\infty}F_2(n,o) = \frac{3}{14} \ .
\eeqn
\end{cor} 

In order to derive the probability $F_j(n,x)$ for arbitrary vertex $x \neq o$, we need the following lemma:

\begin{lemma}
\label{lemmaGH}
For the Sierpinski gasket $SG(n)$ with non-negative integer $n$,
\beqn
G_{0}(n,b_n) & = & \frac{33}{28} \bigl(\frac {3}{5}\bigr)^n - \frac{5}{28} \bigl(\frac {1}{25}\bigr)^n \ , \cr\cr
H_0(n,o) & = & \frac{11}{14} \bigl(\frac {3}{5}\bigr)^n + \frac3{14} \bigl(\frac {1}{25}\bigr)^n \ , \cr\cr
G_{1}(n,b_n) & = & \frac{11}{14} - \frac{2}{7} \bigl(\frac{1}{15}\bigr)^n - \frac{6}{7} \bigl(\frac{3}{5}\bigr)^n + \frac{5}{14} \bigl(\frac{1}{25}\bigr)^n \ , \cr\cr
G_{2}(n,b_n) & = & \frac{3}{14} + \frac{2}{7} \bigl(\frac{1}{15}\bigr)^n - \frac{9}{28} \bigl(\frac{3}{5}\bigr)^n - \frac{5}{28} \bigl(\frac{1}{25}\bigr)^n \ , \cr\cr
H_{1}(n,o) & = & \frac{11}{14} + \frac{3}{14} \bigl(\frac{1}{15}\bigr)^n - \frac{4}{7} \bigl(\frac{3}{5}\bigr)^n - \frac{3}{7} \bigl(\frac{1}{25}\bigr)^n \ , \cr\cr
H_2(n,o) & = & \frac{3}{14} - \frac{3}{14} \bigl(\frac{1}{15}\bigr)^n - \frac{3}{14} \bigl(\frac{3}{5}\bigr)^n + \frac{3}{14} \bigl(\frac{1}{25}\bigr)^n \ .
\eeqn
\end{lemma}

{\sl Proof} \quad  
Denote the vector $Z_0(n)=(G_{0}(n,b_n),H_0(n,o))^T$. By (\ref{iv0}) and (\ref{GH0}), we have $Z_0(0)=(1,1)^T$ and
\[
Z_0(n) = B^nZ_0(n) = Q_BD_B^n Q_B^{-1}Z_0(0) \ ,
\]
where
\beq
B=\left[\begin{array}{cc}
\frac 25 & \frac 3{10} \cr
\frac {6}{25} & \frac 6{25}
\end{array}
\right] \ , \quad
D_B = \left[ \begin{array}{cc}
\frac 35 & 0 \cr
0 & \frac 1{25}
\end{array}
\right] \ , \quad
Q_B = \left[ \begin{array}{cc}
3 & 5 \cr
2 & -6
\end{array}
\right] \ .
\label{BDQ}
\eeq
Then 
\[
Z_0(n) = \left[\begin{array}{cc}
\frac{9}{14} \bigl(\frac{3}{5}\bigr)^n + \frac{5}{14} \bigl(\frac{1}{25}\bigr)^n & \frac{15}{28} \bigl(\frac{3}{5}\bigr)^n - \frac{15}{28} \bigl(\frac{1}{25}\bigr)^n \cr
\frac{3}{7} \bigl(\frac{3}{5}\bigr)^n - \frac{3}{7} \bigl(\frac{1}{25}\bigr)^n & \frac{5}{14} \bigl(\frac{3}{5}\bigr)^n + \frac{9}{14} \bigl(\frac{1}{25}\bigr)^n
\end{array}
\right]Z_0(0) 
= \left[\begin{array}{c}
\frac{33}{28} \bigl(\frac{3}{5}\bigr)^n - \frac{5}{28} \bigl(\frac{1}{25}\bigr)^n \cr
\frac{11}{14} \bigl(\frac{3}{5}\bigr)^n + \frac{3}{14} \bigl(\frac{1}{25}\bigr)^n
\end{array}
\right]
\]
gives $G_0(n,b_n)$ and $H_0(n,o)$.

For the other probabilities, denote the vector $W_j(n)=(G_{j}(n,b_n),H_j(n,o))^T$ with $j=1,2$, and split it into two parts $W_j(n)=W^{(1)}_j(n)+W^{(2)}_j(n)$ \cite{GS}. By (\ref{GHj}), we have
\beq
W^{(1)}_j(n+1) = BW^{(1)}_j(n) \quad \mbox{and} \quad 
W^{(2)}_j(n+1) = BW^{(2)}_j(n)+R_j(n) \ ,
\label{UV}
\eeq
where
\[
R_j(n)=\left[\begin{array}{cc}
& \frac1{10} F_j(n,o) + \frac15 G_{j}(n,o) \cr
& \frac6{25} F_j(n,o) + \frac7{25} G_{j}(n,o) 
\end{array}
\right] \ .
\]
From Theorem \ref{theoremFG}, we know
\[
R_1(n) = \left[\begin{array}{c}
\frac{33}{140} + \frac{13}{420} \bigl(\frac{1}{15}\bigr)^n \cr
\frac{143}{350} + \frac{11}{350} \bigl(\frac{1}{15}\bigr)^n
\end{array}
\right]
\quad \mbox{and} \quad
R_2(n) = \left[\begin{array}{c}
\frac{9}{140} - \frac{13}{420} \bigl(\frac{1}{15}\bigr)^n \cr
\frac{39}{350} - \frac{11}{350} \bigl(\frac{1}{15}\bigr)^n 
\end{array}
\right] \ .
\]
To solve $W_j^{(2)}(n)$, denote
\[
W_1^{(2)}(n) = \left[\begin{array}{c}
a_1 + a_2 \bigl(\frac{1}{15}\bigr)^n \cr
a_3 + a_4 \bigl(\frac{1}{15}\bigr)^n
\end{array}
\right]
\quad \mbox{and} \quad
W^{(2)}_2(n) = \left[\begin{array}{c}
b_1 + b_2 \bigl(\frac{1}{15}\bigr)^n \cr
b_3 + b_4 \bigl(\frac{1}{15}\bigr)^n
\end{array}
\right] \ ,
\]
where the unknown $a_i$ and $b_i$ with $i \in \{1,2,3,4\}$ satisfy the following relations by (\ref{UV}):
\beq
\left\{\begin{array}{lll}
a_1 & = & \frac{33}{140} + \frac{2}{5}a_1 + \frac{3}{10}a_3 \cr
\frac{a_2}{15} & = & \frac{13}{420} + \frac{2}{5}a_2 + \frac{3}{10}a_4 \cr
a_3 & = & \frac{143}{350} + \frac{6}{25} \bigl(a_1+a_3\bigr) \cr
\frac{a_4}{15} & = & \frac{11}{350} + \frac{6}{25} \bigl(a_2+a_4\bigr)
\end{array}
\right. \ , \quad
\left\{\begin{array}{lll}
b_1 & = & \frac{9}{140} + \frac{2}{5}b_1 + \frac{3}{10}b_3 \cr
\frac{b_2}{15} & = & -\frac{13}{420} + \frac{2}{5}b_2 + \frac{3}{10}b_4 \cr
b_3 & = & \frac{39}{350} + \frac{6}{25} \bigl(b_1+b_3\bigr) \cr
\frac{b_4}{15} & = & -\frac{11}{350} + \frac{6}{25} \bigl(b_2+b_4\bigr)
\end{array}
\right. \ .
\label{ab}
\eeq
It is straightforward to solve (\ref{ab}) and obtain
\beq
W_1^{(2)}(n) = \left[\begin{array}{c}
\frac{11}{14} - \frac{2}{7} \bigl(\frac{1}{15}\bigr)^n \cr
\frac{11}{14} + \frac{3}{14} \bigl(\frac{1}{15}\bigr)^n
\end{array}
\right]
\quad \mbox{and} \quad
W_2^{(2)}(n) = \left[\begin{array}{c}
\frac{3}{14} + \frac{2}{7} \bigl(\frac{1}{15}\bigr)^n \cr
\frac{3}{14} - \frac{3}{14} \bigl(\frac{1}{15}\bigr)^n
\end{array}
\right] \ ,
\label{w2}
\eeq
such that $W_1^{(2)}(0)=(1/2,1)^T$ and $W^{(2)}_2(0)=(1/2,0)^T$.

As $W_j(0)=(0,0)^T$ for $j=1,2$, we have $W_1^{(1)}(0)=(-1/2,-1)^T$, $W^{(1)}_2(0)=(-1/2,0)^T$, and $W_j^{(1)}(n)$ can be solved by (\ref{BDQ}), (\ref{UV}) as
\beq
W^{(1)}_1(n) = \left[\begin{array}{cc}
\frac{9}{14} \bigl(\frac{3}{5}\bigr)^n + \frac{5}{14} \bigl(\frac{1}{25}\bigr)^n  
& \frac{15}{28} \bigl(\frac{3}{5}\bigr)^n - \frac{15}{28} \bigl(\frac{1}{25}\bigr)^n \cr
\frac{3}{7} \bigl(\frac{3}{5}\bigr)^n - \frac{3}{7} \bigl(\frac{1}{25}\bigr)^n  
& \frac{5}{14} \bigl(\frac{3}{5}\bigr)^n + \frac{9}{14} \bigl(\frac{1}{25}\bigr)^n
\end{array}
\right]W^{(1)}_1(0) 
= \left[\begin{array}{c}
- \frac{6}{7} \bigl(\frac{3}{5}\bigr)^n + \frac{5}{14} \bigl(\frac{1}{25}\bigr)^n \cr
- \frac{4}{7} \bigl(\frac{3}{5}\bigr)^n - \frac{3}{7} \bigl(\frac{1}{25}\bigr)^n
\end{array}
\right] 
\label{w11}
\eeq
and
\beq
W^{(1)}_2(n) = \left[\begin{array}{cc}
\frac{9}{14} \bigl(\frac{3}{5}\bigr)^n + \frac{5}{14} \bigl(\frac{1}{25}\bigr)^n 
& \frac{15}{28} \bigl(\frac{3}{5}\bigr)^n - \frac{15}{28} \bigl(\frac{1}{25}\bigr)^n \cr
\frac {3}{7} \bigl(\frac{3}{5}\bigr)^n - \frac{3}{7} \bigl(\frac{1}{25}\bigr)^n  
& \frac{5}{14} \bigl(\frac{3}{5}\bigr)^n + \frac{9}{14} \bigl(\frac{1}{25}\bigr)^n
\end{array}
\right]W^{(1)}_2(0) 
= \left[\begin{array}{c}
- \frac{9}{28} \bigl(\frac{3}{5}\bigr)^n - \frac{5}{28} \bigl(\frac{1}{25}\bigr)^n \cr
- \frac {3}{14} \bigl(\frac{3}{5}\bigr)^n + \frac{3}{14} \bigl(\frac{1}{25}\bigr)^n
\end{array} 
\right] \ .
\label{w12}
\eeq
Combining (\ref{w2})-(\ref{w12}), $G_j(n,b_n)$ and $H_j(n,o)$ with $j=1,2$ are solved.
\ $\Box$

\setcounter{equation}{0}
\section{$F_j(n+m+1,x_n)$ with $x \in \{a,b,c\}$ and $n \geq 0$, $m \geq 0$} 
\label{sectionIV}
Consider the Sierpinski gasket $SG(n+m+1)$ with $n \geq 0$, $m \geq 0$. We will derive $F_j(n+1,x_n)$ with $j \in \{1,2,3,4\}$ for the vertex $x_n \in \{a_n, b_n, c_n\}$ first, then $F_j(n+m+1,x_n)$ with arbitrary $m > 0$ in this section. The corresponding $G_j(n+1,x_n)$ and $H_j(n+1,x_n)$ with $j \in \{1,2,3,4\}$ will be used in the next section. Notice that $G_0(n+1,x_n)=H_0(n+1,x_n)=0$ for $x_n \in \{a_n, b_n, c_n\}$ as these vertices are not outmost vertices of $SG(n+1)$.

For the Sierpinski gasket $SG(n+1)$, we know $f_j(n+1,a_n)=f_j(n+1,b_n)=f_j(n+1,c_n)$ and $h_j(n+1,a_n)=h_j(n+1,b_n)=h_j(n+1,c_n)$ with $j \in \{1,2,3,4\}$ because of rotation symmetry. From the definition of $g_j(n,x)$, we have $g_j(n+1,b_n)=g_j(n+1,c_n)$ but they are distinct from $g_j(n+1,a_n)$. Using Figs. \ref{ffig}-\ref{hfig} for the vertex $a_n$ or $b_n$, we obtain the following recursion relations 
\beqn
\left\{\begin{array}{lll}
f_1(n+1,a_n) & = & 2f_1(n,o)g_{0}(n,b_n)f(n) \ , \cr
f_2(n+1,a_n) & = & 2f_2(n,o)g_{0}(n,b_n)f(n) + 2f_1(n,o)\bigl[g_{1}(n,o)+g_{1}(n,b_n)\bigr]f(n) \\
& & + 2f_1(n,o)^2g(n) \ , \cr
f_3(n+1,a_n) & = & 2f_2(n,o)[g_{1}(n,o)+g_{1}(n,b_n)]f(n) + 4f_1(n,o)f_2(n,o)g(n) \\
& & + 2f_1(n,o)[g_{2}(n,o) + g_{2}(n,b_n)]f(n) \ , \cr
f_4(n+1,a_n) & = & 2f_2(n,o)[g_{2}(n,o)+g_{2}(n,b_n)]f(n) + 2f_2(n,o)^2g(n) \ ,
\end{array} \right.
\eeqn
\beqn
\left\{\begin{array}{lll}
g_{1}(n+1,a_n) & = & 2f_1(n,o)g_{0}(n,b_n)g(n) \ , \cr
g_{2}(n+1,a_n) & = & f_{1}(n,o)^2h(n) + g_{1}(n,o)^2f(n) + 4f_1(n,o)g_{1}(n,o)g(n) \\
& & + 2[f_1(n)g_{1}(n,b_n) + f_2(n)g_{0}(n,b_n)]g(n) \ , \cr
g_{3}(n+1,a_n) & = & 2f_2(n,o)f_{1}(n,o)h(n) + 2g_{1}(n,o)g_{2}(n,o)f(n) \\
& & + 4[f_{2}(n,o)g_{1}(n,o) + f_{1}(n,o)g_{2}(n,o)]g(n) \\
& & + 2[f_2(n,o)g_{1}(n,b_n) + f_1(n,o)g_{2}(n,b_n)]g(n) \ , \cr
g_{4}(n+1,a_n) & = & f_2(n,o)^2h(n) + f_2(n,o)[2g_{2}(n,b_n) + 4g_{2}(n,o)]g(n) + g_{2}(n,o)^2f(n) \ ,
\end{array}\right.
\eeqn
\beqn
\left\{\begin{array}{lll}
g_{1}(n+1,b_n) & = & f_1(n,o)h_{0}(n,o)f(n) + 2f_1(n,o)g_{0}(n,b_n)g(n) \\
& & + 2g_{1}(n,o)g_{0}(n,b_n)f(n) \ , \cr
g_{2}(n+1,b_n) & = & [f_2(n,o)h_0(n,o) + f_1(n,o)h_1(n,o)]f(n) \\
& & + 2f_1(n,o)g_{1}(n,o)g(n) + g_{1}(n,o)^2f(n) \\
& & + 2[f_2(n,o)g_{0}(n,b_n) + f_1(n,o)g_{1}(n,b_n)]g(n) \\
& & + 2[g_{2}(n,o)g_{0}(n,b_n) + g_{1}(n,o)g_{1}(n,b_n)]f(n) \ , \cr
g_{3}(n+1,b_n) & = & [f_2(n,o)h_1(n,o) + f_1(n,o)h_2(n,o)]f(n) \\
& & + 2[f_2(n,o)g_{1}(n,o) + f_1(n,o)g_{2}(n,o)]g(n) \\
& & + 2[f_2(n,o)g_{1}(n,b_n) + f_1(n,o)g_{2}(n,b_n)]g(n) \\
& & + 2[g_{2}(n,o)g_{1}(n,b_n) + g_{1}(n,o)g_{2}(n,b_n)]f(n) \\
& & + 2g_{1}(n,o)g_{2}(n,o)f(n) \ , \cr
g_{4}(n+1,b_n) & = & f_2(n,o)h_2(n,o)f(n) + 2g_{2}(n,o)g_{2}(n,b_n)f(n) \\
& & + 2f_2(n,o)[g_{2}(n,o) + g_{2}(n,b_n)]g(n) + g_{2}(n,o)^2f(n) \ ,
\end{array}\right.
\eeqn
and 
\beqn
\left\{\begin{array}{lll}
h_1(n+1,a_n) & = & 4f_1(n,o)h_{0}(n,o)g(n) + 2f_1(n,o)g_{0}(n,b_n)h(n) \\
& & + 4g_{1}(n,o)h_{0}(n,o)f(n) + 8g_{1}(n,o)g_{0}(n,b_n)g(n) \ , \cr
h_2(n+1,a_n) & = & 4[f_2(n,o)h_0(n,o) + f_1(n,o)h_1(n,o)]g(n) \\
& & + 2[f_1(n,o)g_{1}(n,b_n) + f_2(n,o)g_{0}(n,b_n)]h(n) \\
& & + 4[g_{2}(n,o)h_{0}(n,o) + g_{1}(n,o)h_{1}(n,o)]f(n) \\
& & + 2f_1(n,o)g_{1}(n,o)h(n) + 6g_{1}(n,o)^2g(n) \\
& & + 8[g_{2}(n,o)g_{0}(n,b_n) + g_{1}(n,o)g_{1}(n,b_n)]g(n) \ , \cr
h_3(n+1,a_n) & = & 4[f_2(n,o)h_1(n,o) + f_1(n,o)h_2(n,o)]g(n) \\
& & + 2[f_2(n,o)g_{1}(n,b_n) + f_1(n,o)g_{2}(n,b_n)]h(n) \\
& & + 4[g_{2}(n,o)h_{1}(n,o) + g_{1}(n,o)h_{2}(n,o)]f(n) \\
& & + 2[f_2(n,o)g_{1}(n,o) + f_1(n,o)g_2(n,o)]h(n) \\
& & + 8[g_{2}(n,o)g_{1}(n,b_n) + g_{1}(n,o)g_{2}(n,b_n)]g(n) \\
& & + 12g_{1}(n,o)g_{2}(n,o)g(n) \ , \cr
h_4(n+1,a_n) & = & 4f_2(n,o)h_2(n,o)g(n) + 2f_2(n,o)[g_{2}(n,b_n) + g_{2}(n,o)]h(n) \\
& & + 4g_{2}(n,o)h_{2}(n,o)f(n) + 8g_{2}(n,o)g_{2}(n,b_n)g(n) \\
& & + 6g_{2}(n,o)^2g(n) \ .
\end{array}\right.
\eeqn

Using the identity $3g(n)^2=f(n)h(n)$, it follows that
\beqs
\left\{\begin{array}{lll}
F_1(n+1,a_n) & = & \frac{F_1(n,o)G_{0}(n,b_n)}{3} \ , \cr
F_2(n+1,a_n) & = & \frac{F_2(n,o)G_{0}(n,b_n)}{3} + \frac{F_1(n,o)^2}{3} + \frac{F_1(n,o)[G_{1}(n,o) + G_{1}(n,b_n)]}{3} \ , \cr
F_3(n+1,a_n) & = & \frac{F_2(n,o)[G_{1}(n,o) + G_{1}(n,b_n)]}{3} + \frac{2F_1(n,o)F_{2}(n,o)}{3} + \frac{F_1(n,o)[G_{2}(n,o)+G_{2}(n,b_n)]}{3} \ , \cr
F_4(n+1,a_n) & = & \frac{F_2(n,o)[G_{2}(n,o) + G_{2}(n,b_n)]}{3} + \frac{F_2(n,o)^2}{3} \ ,
\end{array}\right.
\label{ivFn}
\eeqs
\beqs
\left\{\begin{array}{lll}
G_{1}(n+1,a_n) & = & \frac{F_1(n,o)G_{0}(n,b_n)}{5} \ , \cr
G_{2}(n+1,a_n) & = & \frac{3F_1(n,o)^2}{10} + \frac{F_1(n,o)[2G_{1}(n,o)+G_1(n,b_n)]}{5} + \frac{F_2(n,o)G_{0}(n,b_n)}{5} + \frac{G_{1}(n,o)^2}{10} \ , \cr
G_{3}(n+1,a_n) & = & \frac{3F_2(n,o)F_1(n,o)}{5} + \frac{F_2(n,o)[2G_{1}(n,o) + G_{1}(n,b_n)]}{5} + \frac{G_{2}(n,o)G_{1}(n,o)}{5} \cr
& & + \frac{F_1(n,o)[2G_{2}(n,o) + G_{2}(n,b_n)]}{5} \ , \cr
G_{4}(n+1,a_n) & = & \frac{3F_2(n,o)^2}{10} + \frac{F_2(n,o)[2G_{2}(n,o) + G_{2}(n,b_n)]}{5} + \frac{G_{2}(n,o)^2}{10} \ ,
\end{array}\right.
\label{ivGan}
\eeqs
\beqs
\left\{\begin{array}{lll}
G_{1}(n+1,b_n) & = & \frac {3F_1(n,o)H_0(n,o)}{10} + \frac{[F_1(n,o) + G_{1}(n,o)]G_{0}(n,b_n)}{5} \ , \cr
G_{2}(n+1,b_n) & = & \frac{3[F_2(n,o)H_0(n,o) + F_1(n,o)H_1(n,o)]}{10} + \frac{G_{1}(n,o)^2}{10} + \frac{F_1(n,o)[G_{1}(n,o) + G_{1}(n,b_n)]}{5} \cr
& & + \frac{F_2(n,o)G_{0}(n,b_n)}{5} + \frac{G_{0}(n,b_n)G_{2}(n,o) + G_{1}(n,o)G_{1}(n,b_n)}{5} \ , \cr
G_{3}(n+1,b_n) & = & \frac{3[F_2(n,o)H_1(n,o) + F_1(n,o)H_2(n,o)]}{10} + \frac{[G_{1}(n,b_n) + G_{1}(n,o)]G_{2}(n,o) + G_{1}(n,o)G_{2}(n,b_n)}{5} \cr
& & + \sum_{r=1}^2 \frac{F_r(n,o)[G_{3-r}(n,o) + G_{3-r}(n,b_n)]}{5} \ , \cr
G_{4}(n+1,b_n) & = & \frac{3F_2(n,o)H_2(n,o)}{10} + \frac{F_2(n,o)[G_{2}(n,o)+G_{2}(n,b_n)] + G_{2}(n,o)G_{2}(n,b_n)}{5} + \frac{G_{2}(n,o)^2}{10} \ ,
\end{array}\right.
\label{ivGbn}
\eeqs
and
\beqs
\left\{\begin{array}{lll}
H_1(n+1,a_n) & = & \frac{3F_1(n,o)[2H_0(n,o) + G_{0}(n,b_n)]}{25} + \frac{6G_{1}(n,o)H_{0}(n,o)}{25} + \frac{4G_{1}(n,o)G_{0}(n,b_n)}{25} \ , \cr
H_2(n+1,a_n) & = & \frac{6[F_2(n,o)H_0(n,o) + F_1(n,o)H_1(n,o)]}{25} + \frac{3F_1(n,o)[G_{1}(n,o) + G_{1}(n,b_n)]}{25} \cr
& & + \frac{3F_2(n,o)G_{0}(n,b_n)}{25} + \frac{4[G_{2}(n,o)G_{0}(n,b_n) + G_{1}(n,o)G_{1}(n,b_n)]}{25} \cr
& & + \frac{3G_{1}(n,o)^2}{25} + \frac{6[H_0(n,o)G_{2}(n,o) + H_1(n,o)G_{1}(n,o)]}{25} \ , \cr
H_3(n+1,a_n) & = & \frac{6[F_2(n,o)H_1(n,o) + F_1(n,o)H_2(n,o)]}{25} + \frac{3F_2(n,o)[G_{1}(n,o) + G_{1}(n,b_n)]}{25} \cr
& & + \frac{3F_1(n,o)[G_{2}(n,o) + G_{2}(n,b_n)]}{25} + \frac{6[G_{2}(n,o)H_1(n,o) + G_{1}(n,o)H_2(n,o)]}{25} \cr
& & + \frac{[4G_{1}(n,b_n) + 6G_{1}(n,o)]G_{2}(n,o)}{25} + \frac{4G_{1}(n,o)G_{2}(n,b_n)}{25} \ , \cr
H_4(n+1,a_n) & = & \frac{6F_2(n,o)H_2(n,o)}{25} + \frac{3F_2(n,o)[G_{2}(n,o) + G_{2}(n,b_n)]}{25} + \frac{6G_{2}(n,o)H_2(n,o)}{25} \cr
& & + \frac{4G_{2}(n,o)G_{2}(n,b_n)}{25} + \frac{3G_{2}(n,o)^2}{25} \ .
\end{array}\right.
\label{ivHn}
\eeqs

Next consider the Sierpinski gasket $SG(n+m+1)$ with $n \geq 0$ and $m > 0$. The left-hand-sides of Figs. \ref{ffig}-\ref{hfig} now represent $SG(n+m+1)$ with positive integer $m$, such that $x_n \in \{a_n,b_n,c_n\}$ locates within the lower-left triangle representing $SG(n+m)$ in the right-hand-sides of the figures. As the vertices are denoted such that $x_{\vec{\gamma}}$ and $\tilde{x}_{\vec{\gamma}}$ are reflection of each other with respect to the extended line connecting $o$ and $c_0$, we have $\tilde{a}_n=b_n$, $\tilde{b}_n=a_n$ and $\tilde{c}_n=c_n$. We obtain the following recursion relations for $j \in \{1,2,3,4\}$:
\beqs
F_j(n+m+1,x_n) & = & \frac{4f_j(n+m,x_n)f(n+m)g(n+m)}{6f(n+m)^2g(n+m)} \cr
& & + \frac{[g_j(n+m,x_n)+g_j(n+m,\tilde{x}_n)]f(n+m)^2}{6f(n+m)^2g(n+m)} \cr
& = & \frac {2F_j(n+m,x_n)}{3} + \frac{G_{j}(n+m,x_n)+G_{j}(n+m,\tilde{x}_n)}{6} \ , 
\label{Fn}
\eeqs
and
\beq
G_{j}(n+m+1,x_n) = \frac{3F_j(n+m,x_n)}{5} + \frac{3G_{j}(n+m,x_n)}{10} + \frac{G_{j}(n+m,\tilde{x}_n)}{10} \ . 
\label{Gn}
\eeq
By symmetry, we know $F_j(n+m,\tilde{x}_n)=F_j(n+m,x_n)$ with $x_n \in \{a_n,b_n,c_n\}$ and $G_j(n+m,\tilde{c}_n)=G_j(n+m,c_n)$ for positive integer $m$.  
Let us define the $3 \times 3$ matrix
\beq
B^\prime_j(n+m+1,n) = \left[\begin{array}{ccc}
F_j(n+m+1,a_n) & G_j(n+m+1,a_n) & G_j(n+m+1,b_n) \\
F_j(n+m+1,b_n) & G_j(n+m+1,b_n) & G_j(n+m+1,a_n) \\
F_j(n+m+1,c_n) & G_j(n+m+1,c_n) & G_j(n+m+1,c_n)
\end{array}
\right] 
\label{Bprime}
\eeq
for non-negative integer $n$, $m$ and $j \in \{1,2,3,4\}$. For $m=0$, $B^\prime_j(n+1,n)$ has been obtained with elements given in (\ref{ivFn})-(\ref{ivGbn}). By (\ref{Fn}) and (\ref{Gn}), we have
\[
B^\prime_j(n+m+1,n) = B^\prime_j(n+m,n) L^\prime \ ,
\]
for any $m \geq 1$, where
\beq
L^\prime = \left[\begin{array}{ccc}
\frac {2}{3} & \frac {3}{5}  & \frac 35 \cr
\frac {1}{6} & \frac {3}{10} & \frac 1{10} \cr
\frac {1}{6} & \frac {1}{10} & \frac 3{10}
\end{array}
\right] \ .
\label{Lprime}
\eeq
We arrive at
\beq
B^\prime_j(n+m+1,n) = B^\prime_j(n+1,n) {L^\prime}^m \quad \mbox{for all} \ m \geq 0, n \geq 0 \ .
\label{BpLpm}
\eeq
Solving $B^\prime_j(n+m+1,n)$ as in the proof of Theorem \ref{theoremFG}, its first column gives 
\beqn
\left[\begin{array}{c}
F_j(n+m+1,a_{n}) \\
F_j(n+m+1,b_{n}) \\
F_j(n+m+1,c_{n})
\end{array} \right] & = & [\frac9{14} + \frac5{14}(\frac 1{15})^m] \left[\begin{array}{c}
F_j(n+1,a_n) \\
F_j(n+1,b_n) \\
F_j(n+1,c_n)
\end{array} \right] \\
& & + \frac{5}{28}[1-(\frac 1{15})^m] \left[\begin{array}{c}
G_j(n+1,a_n) + G_j(n+1,b_n) \\
G_j(n+1,b_n) + G_j(n+1,a_n) \\
2G_j(n+1,c_n)
\end{array}\right] \ ,
\eeqn
and we have the following theorem using (\ref{ivFn})-(\ref{ivGbn}).

\begin{theo} 
\label{theoremF}
For the Sierpinski gasket $SG(n+m+1)$ with non-negative integer $n$ and $m$,
\beqn
\left\{ \begin{array}{l}
F_1(n+m+1,a_n) = \frac{1815}{5488} \left(\frac35\right)^m - \frac{99}{1372} \left(\frac1{25}\right)^m + \frac{55}{16464} \left(\frac1{375}\right)^m \cr
\qquad + \left(\frac 1{15}\right)^m \Bigl \{ - \frac{121}{5488} \left(\frac35\right)^m - \frac{22}{1029}\left(\frac1{25}\right)^m + \frac{185}{49392} \left(\frac1{375}\right)^m \Bigr \} \ , \cr
F_2(n+m+1,a_n) = \frac{121}{196} - \frac{825}{5488} \left(\frac35\right)^m + \frac{171}{1372} \left(\frac1{25}\right)^m - \frac{55}{5488} \left(\frac1{375}\right)^m - \frac{121}{1176} \left(\frac1{15}\right)^m + \frac1{294} \left(\frac1{225}\right)^m \cr
\qquad + \left(\frac 1{15}\right)^m \Bigl \{ \frac{55}{5488} \left(\frac35\right)^m + \frac{38}{1029} \left(\frac1{25}\right)^m - \frac{185}{16464} \left(\frac1{375}\right)^m - \frac{143}{3528} \left(\frac1{15}\right)^m + \frac{11}{2646} \left(\frac1{225}\right)^m \Bigr \} \ , \cr
F_3(n+m+1,a_n) = \frac{33}{98} - \frac{855}{5488} \left(\frac35\right)^m - \frac{45}{1372} \left(\frac1{25}\right)^m + \frac{55}{5488} \left(\frac1{375}\right)^m + \frac{11}{147} \left(\frac1{15}\right)^m - \frac1{147} \left(\frac1{225}\right)^m \cr
\qquad + \left(\frac 1{15}\right)^m \Bigl \{ \frac{57}{5488} \left(\frac35\right)^m - \frac{10}{1029} \left(\frac1{25}\right)^m + \frac{185}{16464} \left(\frac1{375}\right)^m + \frac{13}{441} \left(\frac1{15}\right)^m - \frac{11}{1323} \left(\frac1{225}\right)^m \Bigr \} \ , \cr
F_4(n+m+1,a_n) = \frac{9}{196} - \frac{135}{5488} \left(\frac35\right)^m - \frac{27}{1372} \left(\frac1{25}\right)^m - \frac{55}{16464} \left(\frac1{375}\right)^m + \frac{11}{392} \left(\frac1{15}\right)^m + \frac1{294} \left(\frac1{225}\right)^m \cr
\qquad + \left(\frac 1{15}\right)^m \Bigl \{ \frac{9}{5488} \left(\frac35\right)^m - \frac{2}{343} \left(\frac1{25}\right)^m - \frac{185}{49392} \left(\frac1{375}\right)^m + \frac{13}{1176} \left(\frac1{15}\right)^m + \frac{11}{2646} \left(\frac1{225}\right)^m \Bigr \} \ , 
\end{array} \right.
\eeqn
\beqn
\left\{ \begin{array}{l}
F_1(n+m+1,c_n) = \frac{1089}{2744} \left(\frac35\right)^m - \frac{22}{343} \left(\frac1{25}\right)^m + \frac{5}{8232} \left(\frac1{375}\right)^m \cr
\qquad + \left(\frac 1{15}\right)^m \Bigl \{ - \frac{121}{1372} \left(\frac35\right)^m - \frac{121}{4116} \left(\frac1{25}\right)^m + \frac{20}{3087} \left(\frac1{375}\right)^m \Bigr \} \ , \cr
F_2(n+m+1,c_n) = \frac{121}{196} - \frac{495}{2744} \left(\frac35\right)^m +  \frac{38}{343} \left(\frac1{25}\right)^m - \frac{5}{2744} \left(\frac1{375}\right)^m - \frac{55}{588} \left(\frac1{15}\right)^m \cr
\qquad + \left(\frac 1{15}\right)^m \Bigl \{ \frac{55}{1372} \left(\frac35\right)^m +  \frac{209}{4116} \left(\frac1{25}\right)^m - \frac{20}{1029} \left(\frac1{375}\right)^m - \frac{22}{441} \left(\frac1{15}\right)^m + \frac{10}{1323} \left(\frac1{225}\right)^m \Bigr \} \ , \cr
F_3(n+m+1,c_n) = \frac{33}{98} - \frac{513}{2744} \left(\frac35\right)^m - \frac{10}{343} \left(\frac1{25}\right)^m + \frac{5}{2744}\left(\frac1{375}\right)^m + \frac{10}{147} \left(\frac1{15}\right)^m \cr
\qquad + \left(\frac 1{15}\right)^m \Bigl \{ \frac{57}{1372} \left(\frac35\right)^m - \frac{55}{4116} \left(\frac1{25}\right)^m + \frac{20}{1029} \left(\frac1{375}\right)^m + \frac{16}{441} \left(\frac1{15}\right)^m - \frac{20}{1323} \left(\frac1{225}\right)^m \Bigr \} \ , \cr
F_4(n+m+1,c_n) = \frac{9}{196} - \frac{81}{2744} \left(\frac35\right)^m - \frac{6}{343} \left(\frac1{25}\right)^m - \frac{5}{8232} \left(\frac1{375}\right)^m + \frac{5}{196} \left(\frac1{15}\right)^m \cr
\qquad + \left(\frac 1{15}\right)^m \Bigl \{ \frac{9}{1372} \left(\frac35\right)^m - \frac{11}{1372} \left(\frac1{25}\right)^m - \frac{20}{3087} \left(\frac1{375}\right)^m + \frac{2}{147} \left(\frac1{15}\right)^m + \frac{10}{1323} \left(\frac1{225}\right)^m \Bigr \} \ .
\end{array} \right.
\eeqn
\end{theo}

\begin{cor}
\label{corollaryac} 
For the Sierpinski gasket $SG(n+m+1)$ with non-negative integer $n$ and $m$, the limiting probabilities are
\beqn
\begin{array}{cc}
\lim_{m \rightarrow \infty}F_1(n+m+1,x_n) = 0 \ , \qquad &
\lim_{m \rightarrow \infty}F_2(n+m+1,x_n) = \frac{121}{196} \ , \cr
\lim_{m \rightarrow \infty}F_3(n+m+1,x_n) = \frac{33}{98} \ , \qquad &
\lim_{m \rightarrow \infty}F_4(n+m+1,x_n) = \frac{9}{196} \ ,
\end{array}
\eeqn
where the vertex $x_n$ can be either $a_n$, $b_n$ or $c_n$.
\end{cor}

It is intriguing to notice that in Theorem \ref{theoremF}, $F_j(n+m+1,a_n)$ are distinct from $F_j(n+m+1,c_n)$ with $j \in \{1,2,3,4\}$, while they have the same value in the infinite $m$ limit.

\setcounter{equation}{0}
\section{$F_j(n+m,x)$ for general $x_{\vec{\gamma}} \in V(SG(n))$ with $n \ge 0$, $m \ge 0$} 
\label{sectionV}
Consider the Sierpinski gasket $SG(n+m)$ with $n \geq 0$, $m \geq 0$. We will derive in this section $F_j(n+m,x)$ with $j \in \{1,2,3,4\}$ for the general vertex $x_{\vec{\gamma}} \in V(SG(n))$ that has not been considered in previous sections.
For the vertices inside the triangle with outmost vertices $a_{n-1}$, $a_n$ and $c_{n-1}$, let us append subscripts in the notation such that $\vec{\gamma}_{n,s} = (\gamma_1=n-1,\gamma_2,\cdots,\gamma_s)$ with $1\leq s\leq n$ and $\gamma_k \in \{0,1,2\}$ for $k \in \{2,3,...,s\}$. The results obtained in section \ref{sectionIV} correspond to the vertex with $s=1$ and $n \ge 1$, and we will tackle the vertex with $s>1$ here. Similar to the definition of the vertex $\tilde{x}_{\vec{\gamma}_{n,s}}$, let us define the vertex $\hat{x}_{\vec{\gamma}_{n,s}}$ as the reflection of $x_{\vec{\gamma}_{n,s}}$ with respect to the line connecting $a_n$ and $b_{n-1}$. By definition, we have 
\[
\tilde{\tilde{x}}_{\vec{\gamma}_{n,s}} = x_{\vec{\gamma}_{n,s}} \ , \hat{\hat{x}}_{\vec{\gamma}_{n,s}} = x_{\vec{\gamma}_{n,s}} \ , 
\]
where $x$ can be either $a$, $b$ or $c$, and
\[
F_j(n,x_{\vec{\gamma}_{n,s}}) = F_j(n,\tilde{x}_{\vec{\gamma}_{n,s}}) = F_j(n,\hat{x}_{\vec{\gamma}_{n,s}}) \ , \qquad 
H_j(n,x_{\vec{\gamma}_{n,s}}) = H_j(n,\tilde{x}_{\vec{\gamma}_{n,s}}) = H_j(n,\hat{x}_{\vec{\gamma}_{n,s}}) \ ,
\]
due to the symmetry of $SG(n)$. For $m \geq 0$, define the $3\times 5$ matrix 
\beqn
\lefteqn{B_j(n+m,\vec{\gamma}_{n,s})} \cr
& = & \left[\begin{array}{ccccc}
F_j(n+m,a_{\vec{\gamma}_{n,s}}) & G_j(n+m,a_{\vec{\gamma}_{n,s}}) & G_j(n+m,\tilde{a}_{\vec{\gamma}_{n,s}}) & G_j(n+m,\hat{a}_{\vec{\gamma}_{n,s}}) & H_j(n+m,a_{\vec{\gamma}_{n,s}}) \\ 
F_j(n+m,b_{\vec{\gamma}_{n,s}}) & G_j(n+m,b_{\vec{\gamma}_{n,s}}) & G_j(n+m,\tilde{b}_{\vec{\gamma}_{n,s}}) & G_j(n+m,\hat{b}_{\vec{\gamma}_{n,s}}) & H_j(n+m,b_{\vec{\gamma}_{n,s}}) \\
F_j(n+m,c_{\vec{\gamma}_{n,s}}) & G_j(n+m,c_{\vec{\gamma}_{n,s}}) & G_j(n+m,\tilde{c}_{\vec{\gamma}_{n,s}}) & G_j(n+m,\hat{c}_{\vec{\gamma}_{n,s}}) & H_j(n+m,c_{\vec{\gamma}_{n,s}})
\end{array}
\right] \ . \cr & &
%\label{B}
\eeqn
This is a generalization of $B^\prime_j(n+m,n)$ in (\ref{Bprime}), which corresponds to the case with $s=1$. By an argument similar to that of (\ref{BpLpm}), we have
\beq
B_j(n+m,\gamma_{n,s}) = B_j(n,\gamma_{n,s})L^m 
\label{BLm}
\eeq
for $m \geq 0$, where the $5 \times 5$ matrix
\[
L = \left[\begin{array}{ccccc}
\frac{2}{3} & \frac{3}{5}  & \frac35    & \frac35    & \frac{6}{25} \cr
\frac{1}{6} & \frac{3}{10} & \frac1{10} & 0          & \frac{7}{50} \cr
\frac{1}{6} & \frac{1}{10} & \frac3{10} & \frac1{10} & \frac7{50}   \cr
0           & 0            & 0          & \frac3{10} & \frac6{25}   \cr
0           & 0            & 0          & 0          & \frac6{25}
\end{array} 
\right]
%\label{L}
\]
is the generalization of $L^\prime$ in (\ref{Lprime}). It follows that the determination of $B_j(n,\vec{\gamma}_{n,s})$ for $s > 1$ will be sufficient.

Let us first consider the vertices with $s=2$ and $\gamma_2=1$, namely, $\vec{\gamma}_{n,2}=(n-1,1)$ with $n=2,3,...$. We obtain the following equations (cf. Figs. \ref{ffig}-\ref{hfig}):  

\beqs
\left\{ \begin{array}{lll}
F_j(n,x_{(n-1,1)}) & = & \frac{2F_j(n-1,x_{n-2})}{3} + \frac{G_j(n-1,x_{n-2})}{6} + \frac{G_j(n-1,\hat{x}_{n-2})}{6} \ , \cr
G_j(n,x_{(n-1,1)}) & = & \frac{3F_j(n-1,x_{n-2})}{5} + \frac{3G_j(n-1,x_{n-2})}{10} + \frac{G_j(n-1,\hat{x}_{n-2})}{10} \ , \cr
G_j(n,\tilde{x}_{(n-1,1)}) & = & \frac{3H_j(n-1,x_{n-2})}{10} + \frac{F_j(n-1,x_{n-2})}{10} + \frac{2G_j(n-1,\tilde{x}_{n-2})}{5} \cr
& & \quad + \frac{G_j(n-1,x_{n-2})}{10} + \frac{G_j(n-1,\hat{x}_{n-2})}{10} \ , \cr
G_j(n,\hat{x}_{(n-1,1)}) & = & \frac{3F_j(n-1,x_{n-2})}{5} + \frac{3G_j(n-1,\hat {x}_{n-2})}{10} + \frac{G_j(n-1,x_{n-2})}{10} \ , \cr
H_j(n+1,x_{(n-1,1)}) & = & \frac{6F_j(n-1,x_{n-2})}{25} + \frac{7G_j(n-1,x_{n-2})}{50} + \frac{6G_j(n-1,\tilde{x}_{n-2})}{25} \cr
& & + \frac{7G_j(n-1,\hat{x}_{n-2})}{50} + \frac{6H_j(n-1,x_{n-2})}{25} \ ,
\end{array} \right.
\label{FGHnk}
\eeqs
where $x$ can be either $a$, $b$ or $c$. Define the $5 \times 5$ matrix
\beq
R=\left[\begin{array}{ccccc}
\frac23 & \frac35    & \frac{1}{10} & \frac{3}{5}  & \frac{6}{25} \cr
\frac16 & \frac3{10} & \frac{1}{10} & \frac{1}{10} & \frac{7}{50} \cr
0       & 0          & \frac25      & 0            & \frac6{25}   \cr
\frac16 & \frac1{10} & \frac{1}{10} & \frac{3}{10} & \frac{7}{50} \cr
0       & 0          & \frac{3}{10} & 0            & \frac 6{25}
\end{array}
\right] , 
\label{R}
\eeq
then (\ref{FGHnk}) is equivalent to
\beq
B_j\bigl(n,(n-1,1)\bigr) = B_j\bigl(n-1,n-2\bigr)R \ .
\label{RBr} 
\eeq
For general $m \geq 0$, we have the following formula combining (\ref{BLm}) and (\ref{RBr}):
\beq
B_j\bigl(n+m,(n-1,1)\bigr) = B_j\Bigr(n,(n-1,1)\Bigr)L^m = B_j\bigl(n-1,n-2\bigr)RL^m \ .
\label{LBCBk}
\eeq
As $F_j(n+m,x_{n-1,1}) = F_j(n+m,\tilde{x}_{n-1,1})$ for $x=a,b,c$, the first column of the matrix in (\ref{LBCBk}) gives all $F_j(n+m,x_{\vec{\gamma}_{n,2}})$ in terms of the quantities for $x_{n-2}$. 

\begin{propo}
\label{propositionF}
For the Sierpinski gasket $SG(n+m)$ with $n \geq 2$, $m \geq 0$,
\[
\left[\begin{array}{c}
F_j(n+m,a_{n-1,1}) \\
F_j(n+m,b_{n-1,1}) \\
F_j(n+m,c_{n-1,1})
\end{array}\right] = \left[\begin{array}{c}
F_j(n+m,\tilde{a}_{n-1,1}) \\
F_j(n+m,\tilde{b}_{n-1,1}) \\
F_j(n+m,\tilde{c}_{n-1,1})
\end{array}\right] = B_j \bigl(n-1,n-2\bigr) R L^m e_1 \ , 
\]
where
\[
e_1=(1,0,0,0,0)^T 
\]
and $j\in\{1,2,3,4\}$.
\end{propo}

Move on to the general vertex $x_{\vec{\gamma}_{n+1,s}}$ inside the triangle with outmost vertices $a_n$, $a_{n+1}$ and $c_n$ for the Sierpinski gasket $SG(n+1)$, where $\vec{\gamma}_{n+1,s} = (n,1,\gamma_3,...,\gamma_s)$ with $\gamma_k \in \{0,1,2\}$, $k=3,4,..,s$ and $3 \leq s \leq n+1$. As $\gamma_3$ can take three possible values, let us discuss them separately.

First consider the case with $\gamma_3=1$. The vertex $x_{\vec{\gamma}_{n+1,s}} = x_{(n,1,1,\gamma_4,...\gamma_s)}$ is located inside the triangle with outmost vertices $a_{n,1}$, $a_{n+1}$ and $c_{n,1}$. Associate with this $x_{\vec{\gamma}_{n+1,s}}$ a vertex $x_{\vec{\gamma}^1_{n,s-1}}$, where $\vec{\gamma}^1_{n,s-1} = (n-1,1,\gamma_4,...,\gamma_s)$ has $s-1$ component. That is, $\vec{\gamma}^1_{n,s-1}$ is obtained from $\vec{\gamma}_{n+1,s}$ by taking out $\gamma_3=1$ and replacing $\gamma_1=n$ by $n-1$. It can be seen that this vertex $x_{\vec{\gamma}^1_{n,s-1}}$ is located inside the triangle with outmost vertices $a_{n-1}$, $a_n$ and $c_{n-1}$. Moreover, $x_{\vec{\gamma}^1_{n,s-1}}$ can be reached from $x_{\vec{\gamma}_{n+1,s}}$ by a horizontal translation with the distance from $a_n$ to $o$. Particularly, if $s=3$ such that $\vec{\gamma}_{n+1,s}=(n,1,1)$, then $\vec{\gamma}^1_{n,s-1}=(n-1,1)$. By the method obtaining (\ref{RBr}), we have
\beq
B_j\Bigl(n+1,\vec{\gamma}_{n+1,s}\Bigr) = B_j\Bigl(n,\vec{\gamma}^1_{n,s-1}\Bigr) R 
\label{gamma1}
\eeq
if $\gamma_3=1$.

Now consider the case with $\gamma_3=2$. The vertex $x_{\vec{\gamma}_{n+1,s}} = x_{(n,1,2,\gamma_4,...\gamma_s)}$ is located inside the triangle with outmost vertices $b_{n,1}$, $c_{n,1}$ and $c_n$. Associate with this $x_{\vec{\gamma}_{n+1,s}}$ a vertex $\tilde{y}_{\vec{\gamma}^2_{n,s-1}}$ where $\vec{\gamma}^2_{n,s-1} = (n-1,1,\gamma_4^2,...,\gamma^2_s))$ has $s-1$ component. Here $y=b$ when $x=a$ and vice versa, and $y=c$ when $x=c$. Namely, $y$ is related to $x$ with three possibilities: $(x,y)=(a,b)$, $(b,a)$ and $(c,c)$. Similarly, $\gamma^2_k$ is related to $\gamma_k$ with three possibilities: $(\gamma_k,\gamma^2_k)=(1,2)$, $(2,1)$ and $(0,0)$, where $k=4,...,s$. Again, $\tilde{y}_{\vec{\gamma}^2_{n,s-1}}$ can be reached from $x_{\vec{\gamma}_{n+1,s}}$ by a horizontal translation with the distance from $a_n$ to $o$. We have
\beq
B_j\Bigl(n+1,\vec{\gamma}_{n+1,s}\Bigr) = \tilde{B}_j \Bigl(n,\vec{\gamma}^2_{n,s-1}\Bigr) R \ ,
\label{BBtilde}
\eeq
where  
\beq
\tilde{B}_j\Bigl(n,\vec{\gamma}^2_{n,s-1}\Bigr)
= \left[\begin{array}{ccccc}
F_j(n,\tilde{b}_{\vec{\gamma}^2_{n,s-1}}) & G_j(n,\tilde{b}_{\vec{\gamma}^2_{n,s-1}}) & G_j(n,\tilde{\tilde{b}}_{\vec{\gamma}^2_{n,s-1}}) & G_j(n,\hat{\tilde{b}}_{\vec{\gamma}^2_{n,s-1}}) & H_j(n,\tilde{b}_{\vec{\gamma}^2_{n,s-1}}) \\
F_j(n,\tilde{a}_{\vec{\gamma}^2_{n,s-1}}) & G_j(n,\tilde{a}_{\vec{\gamma}^2_{n,s-1}}) & G_j(n,\tilde{\tilde{a}}_{\vec{\gamma}^2_{n,s-1}}) & G_j(n,\hat{\tilde{a}}_{\vec{\gamma}^2_{n,s-1}}) & H_j(n,\tilde{a}_{\vec{\gamma}^2_{n,s-1}}) \\
F_j(n,\tilde{c}_{\vec{\gamma}^2_{n,s-1}}) & G_j(n,\tilde{c}_{\vec{\gamma}^2_{n,s-1}}) & G_j(n,\tilde{\tilde{c}}_{\vec{\gamma}^2_{n,s-1}}) & G_j(n,\hat{\tilde{c}}_{\vec{\gamma}^2_{n,s-1}}) & H_j(n,\tilde{c}_{\vec{\gamma}^2_{n,s-1}})
\end{array}
\right] \! . 
\label{Btilde}
\eeq
By symmetry, the columns in (\ref{Btilde}) can be replaced as
\[
\left[\begin{array}{c}
F_j(n,\tilde{b}_{\vec{\gamma}^2_{n,s-1}}) \\
F_j(n,\tilde{a}_{\vec{\gamma}^2_{n,s-1}}) \\
F_j(n,\tilde{c}_{\vec{\gamma}^2_{n,s-1}})
\end{array}
\right]
= \left[\begin{array}{c}
F_j(n,b_{\vec{\gamma}^2_{n,s-1}}) \\
F_j(n,a_{\vec{\gamma}^2_{n,s-1}}) \\
F_j(n,c_{\vec{\gamma}^2_{n,s-1}})
\end{array}
\right] , \quad \left[\begin{array}{c}
H_j(n,\tilde{b}_{\vec{\gamma}^2_{n,s-1}}) \\
H_j(n,\tilde{a}_{\vec{\gamma}^2_{n,s-1}}) \\
H_j(n,\tilde{c}_{\vec{\gamma}^2_{n,s-1}})
\end{array}\right]
 = \left[\begin{array}{c}
H_j(n,b_{\vec{\gamma}^2_{n,s-1}}) \\
H_j(n,a_{\vec{\gamma}^2_{n,s-1}}) \\
H_j(n,c_{\vec{\gamma}^2_{n,s-1}})
\end{array}
\right] ,
\]
and
\[
\left[\begin{array}{c}
G_j(n,\tilde{\tilde{b}}_{\vec{\gamma}^2_{n,s-1}}) \\
G_j(n,\tilde{\tilde{a}}_{\vec{\gamma}^2_{n,s-1}}) \\
G_j(n,\tilde{\tilde{c}}_{\vec{\gamma}^2_{n,s-1}})
\end{array}
\right]
 = \left[\begin{array}{c}
G_j(n,b_{\vec{\gamma}^2_{n,s-1}}) \\
G_j(n,a_{\vec{\gamma}^2_{n,s-1}}) \\
G_j(n,c_{\vec{\gamma}^2_{n,s-1}})
\end{array}
\right] , \quad \left[\begin{array}{c}
G_j(n,\hat{\tilde{b}}_{\vec{\gamma}^2_{n,s-1}}) \\
G_j(n,\hat{\tilde{a}}_{\vec{\gamma}^2_{n,s-1}}) \\
G_j(n,\hat{\tilde{c}}_{\vec{\gamma}^2_{n,s-1}})
\end{array}\right]
 = \left[\begin{array}{c}
G_j(n,\hat{b}_{\vec{\gamma}^2_{n,s-1}}) \\
G_j(n,\hat{a}_{\vec{\gamma}^2_{n,s-1}}) \\
G_j(n,\hat{c}_{\vec{\gamma}^2_{n,s-1}})
\end{array}
\right] ,
\]
so that
\[
\tilde{B}_j\bigl(n,\vec{\gamma}^2_{n,s-1}\bigr) = B_j\bigl(n,\vec{\gamma}^2_{n,s-1}\bigr)E_2 \ ,
\]
where
\[
E_2=\left[\begin{array}{ccccc}
1 & 0 & 0 & 0 & 0 \\
0 & 0 & 1 & 0 & 0 \\
0 & 1 & 0 & 0 & 0 \\
0 & 0 & 0 & 1 & 0 \\
0 & 0 & 0 & 0 & 1
\end{array}
\right].
\]
(\ref{BBtilde}) can be rewritten as
\beq
B_j(n+1,\vec{\gamma}_{n+1,s}) = B_j(n,\vec{\gamma}^2_{n,s-1})E_2R
\label{gamma2}
\eeq
if $\gamma_3=2$.

Finally consider the case with $\gamma_3=0$. The vertex $x_{\vec{\gamma}_{n+1,s}} = x_{(n,1,0,\gamma_4,...\gamma_s)}$ is located inside the triangle with outmost vertices $a_n$, $a_{n,1}$ and $b_{n,1}$. Associate with this $x_{\vec{\gamma}_{n+1,s}}$ a vertex $\bar{z}_{\vec{\gamma}^0_{n,s-1}}$ where $\vec{\gamma}^0_{n,s-1} = (n-1,1,\gamma^0_4,...,\gamma^0_s)$ has $s-1$ component. Here $z=c$ when $x=b$ and vice versa, and $z=a$ when $x=a$. Namely, $z$ is related to $x$ with three possibilities: $(x,z)=(a,a)$, $(b,c)$ and $(c,b)$. Similarly, $\gamma^0_k$ is related to $\gamma_k$ with three possibilities: $(\gamma_k,\gamma^0_k)=(1,0)$, $(0,1)$ and $(2,2)$, where $k=4,...,s$.
We use the notation such that the vertex $\bar{z}_{\vec{\gamma}^0_{n,s-1}}$ is the reflection of the vertex $x_{\vec{\gamma}_{n,s-1}}$ with respect to the line connecting $a_{n-1}$ and $b_n$. It can be seen that $\bar{z}_{\vec{\gamma}^0_{n,s-1}}$ can be reached from $x_{\vec{\gamma}_{n+1,s}}$ by a horizontal translation with the distance from $a_n$ to $o$. We have
\beq
B_j(n+1,\vec{\gamma}_{n+1,s}) = \bar{B}_j(n,\vec{\gamma}^0_{n,s-1}) R \ ,
\label{BBbar}
\eeq
where 
\beqn
\bar{B}_j\Bigl(n,\vec{\gamma}^0_{n,s-1}\Bigr)
& = & \left[\begin{array}{ccccc}
F_j(n,\bar{a}_{\vec{\gamma}^0_{n,s-1}}) & G_j(n,\bar{a}_{\vec{\gamma}^0_{n,s-1}}) & G_j(n,\tilde{\bar{a}}_{\vec{\gamma}^0_{n,s-1}}) & G_j(n,\hat{\bar{a}}_{\vec{\gamma}^0_{n,s-1}}) & H_j(n,\bar{a}_{\vec{\gamma}^0_{n,s-1}}) \\
F_j(n,\bar{c}_{\vec{\gamma}^0_{n,s-1}}) & G_j(n,\bar{c}_{\vec{\gamma}^0_{n,s-1}}) & G_j(n,\tilde{\bar{c}}_{\vec{\gamma}^0_{n,s-1}}) & G_j(n,\hat{\bar{c}}_{\vec{\gamma}^0_{n,s-1}}) & H_j(n,\bar{c}_{\vec{\gamma}^0_{n,s-1}}) \\
F_j(n,\bar{b}_{\vec{\gamma}^0_{n,s-1}}) & G_j(n,\bar{b}_{\vec{\gamma}^0_{n,s-1}}) & G_j(n,\tilde{\bar{b}}_{\vec{\gamma}^0_{n,s-1}}) & G_j(n,\hat{\bar{b}}_{\vec{\gamma}^0_{n,s-1}}) & H_j(n,\bar{b}_{\vec{\gamma}^0_{n,s-1}})
\end{array}
\right] \\
& = & \left[\begin{array}{ccccc}
F_j(n,a_{\vec{\gamma}^0_{n,s-1}}) & G_j(n,a_{\vec{\gamma}^0_{n,s-1}}) & G_j(n,\hat{a}_{\vec{\gamma}^0_{n,s-1}}) & G_j(n,\tilde{a}_{\vec{\gamma}^0_{n,s-1}}) & H_j(n,a_{\vec{\gamma}^0_{n,s-1}}) \\
F_j(n,c_{\vec{\gamma}^0_{n,s-1}}) & G_j(n,c_{\vec{\gamma}^0_{n,s-1}}) & G_j(n,\hat{c}_{\vec{\gamma}^0_{n,s-1}}) & G_j(n,\tilde{c}_{\vec{\gamma}^0_{n,s-1}}) & H_j(n,c_{\vec{\gamma}^0_{n,s-1}}) \\
F_j(n,b_{\vec{\gamma}^0_{n,s-1}}) & G_j(n,b_{\vec{\gamma}^0_{n,s-1}}) & G_j(n,\hat{b}_{\vec{\gamma}^0_{n,s-1}}) & G_j(n,\tilde{b}_{\vec{\gamma}^0_{n,s-1}}) & H_j(n,b_{\vec{\gamma}^0_{n,s-1}})
\end{array}
\right] 
\eeqn
by symmetry, so that
\[
\bar{B}_j \bigl(n,\vec{\gamma}^0_{n,s-1}\bigr) = B_j \bigl(n,{\vec{\gamma}^0_{n,s-1}}\bigr) E_0 \ ,
\]
where
\[
E_0=\left[\begin{array}{ccccc}
1 & 0 & 0 & 0 & 0 \\
0 & 1 & 0 & 0 & 0 \\
0 & 0 & 0 & 1 & 0 \\
0 & 0 & 1 & 0 & 0 \\
0 & 0 & 0 & 0 & 1
\end{array}
\right] \ .
\]
(\ref{BBbar}) can be rewritten as 
\beq
B_j \bigl(n+1,\vec{\gamma}_{n+1,s}\bigr) = B_j \bigl(n,\vec{\gamma}^0_{n,s-1}\bigr) E_0 R 
\label{gamma0}
\eeq
if $\gamma_3=0$.

Denote $E_1=I_{5\times 5}$ as the identity matrix.  
(\ref{gamma1}), (\ref{gamma2}) and (\ref{gamma0}) can be combined to give
\beq
B_j \bigl(n+1,\vec{\gamma}_{n+1,s}\bigr) = B_j \bigl(n,\vec{\gamma}^{\gamma_3}_{n,s-1}\bigr) E_{\gamma_3} R \ ,
\label{BBER}
\eeq
where $\gamma_3 \in \{0,1,2\}$. As $F_j(n+1,x_{\vec{\gamma}_{n+1,s}}) = F_j(n+1,\tilde{x}_{\vec{\gamma}_{n+1,s}})$ for $x=a,b,c$, the first column of the matrix in (\ref{BBER}) gives $F_j(n+1,x_{\vec{\gamma}_{n+1,s}})$ for any vertex $x_{\vec{\gamma}_{n+1,s}}$ in terms of the quantities for $x_{\vec{\gamma}_{n,s-1}}$. 

\begin{propo}
\label{propositionFF}
For the Sierpinski gasket $SG(n+1)$ with $n \geq 2$, consider the vertex $x_{\vec{\gamma}_{n+1,s}}$ where $\vec{\gamma}_{n+1,s}=(n,1,\gamma_3,...,\gamma_s)$ with $3 \leq s \leq n+1$ and $\gamma_k \in \{0,1,2\}$ for $k \in \{3,4,..,s\}$.
\[
\left[\begin{array}{c}
F_j(n+1,a_{\vec{\gamma}_{n+1,s}}) \\
F_j(n+1,b_{\vec{\gamma}_{n+1,s}}) \\
F_j(n+1,c_{\vec{\gamma}_{n+1,s}})
\end{array} \right] 
= B_j (n,\vec{\gamma}^{\gamma_3}_{n,s-1}) E_{\gamma_3} R e_1 \ .
\]
\end{propo}

Using Theorems \ref{theoremFG}, \ref{theoremF} and Propositions \ref{propositionF}, \ref{propositionFF} repeatedly, $F_j(n+1,x_{\vec{\gamma}_{n+1,s}})$ for all the vertices of $SG(n+1)$ can be obtained. 

\setcounter{equation}{0}
\section{Summation and average of $F_j(n,x)$ over all the vertices of $SG(n)$} 
\label{sectionVI}

It is worthwhile to derive the summation of $F_j(n,x)$ over all the vertices $x$ of $SG(n)$, defined as
\[
\Phi_j(n) = \sum_{x \in V(SG(n))} F_j(n,x) \ ,
\]
and the average of $F_j(n,x)$ over all the vertices, defined as
\[
\phi_j(n) = \frac{\Phi_j(n)}{v(SG(n))} = \frac{\Phi_j(n)}{\frac {3}{2}(3^n+1)} \ .
%\label{phi}
\]
It is clear that for any non-negative integer $n$,
\[
\sum_{j=1}^4 \phi_j(n) = 1 \ .
\]

For the vertex $x_{\vec{\gamma}_{n,s}}$ with $s=1$, i.e. $a_m$, $b_m$ and $c_m$, define their sum
\[
X_j(n,m) = F_j(n,a_m)+F_j(n,b_m)+F_j(n,c_m) \ .
\]
Similarly for the vertex with $s=2$, define
\[
Y_j(n,m') = F_j(n,a_{m',1}) + F_j(n,b_{m',1}) + F_j(n,c_{m',1}) \ ,
\]
where $m \geq 0$, $m' \geq 1$ and $n$ is larger than $m$ and $m'$. By (\ref{BLm}), we have
\beq
X_j(n,m) = (1,1,1) B_j(n,m) e_1 = (1,1,1) B_j(m+1,m) L^{n-m-1} e_1 \ , 
\label{Xnm}
\eeq
\beqs
Y_j(n,m') & = & (1,1,1) B_j \bigl(n,(m',1)\bigr) e_1 = (1,1,1) B_j(m'+1,(m',1)) L^{n-m'-1} e_1 \cr
& = & (1,1,1) B_j(m',m'-1) R L^{n-m'-1} e_1 \ .
\label{Ynm}
\eeqs
The first few $\Phi_j(n)$ are
\beqn
\Phi_j(0) & = & 3F_j(0,o) \ , \cr
\Phi_j(1) & = & 3F_j(1,o) + X_j(1,0) = 3F_j(1,o) + 2F_j(1,a_0) + F_j(1,c_0) \ ,
\eeqn
and
\beqn
\Phi_j(2) & = & 3F_j(2,o) + X_j(2,0) + X_j(2,1) + 2Y_j(2,1) \cr
& = & 3F_j(2,o) + (1,1,1) \Bigl\{B_j(1,0)L + B_j(2,1) + 2B_j(1,0)R\Bigr\}e_1 \ .
\eeqn
The corresponding values for $j \in \{1,2,3,4\}$ are
\beqn
& & \Phi_1(0) = \frac23 \ , \qquad \Phi_2(0) = \frac13 \ , \qquad \Phi_3(0) = \Phi_4(0) = 0 \ , \cr\cr
& & \Phi_1(1) = \frac12 \ , \qquad \Phi_2(1) = \frac{19}{54} \ , \qquad \Phi_3(1) = \frac7{54} \ , \qquad \Phi_4(1) = \frac1{54} \ ,  \cr\cr
& & \Phi_1(2) = \frac{163}{450} \ , \qquad \Phi_2(2) = \frac{5257}{12150} \ , \qquad \Phi_3(2) = \frac{2203}{12150} \ , \qquad \Phi_4(2) = \frac{289}{12150} \ .
\eeqn
For $n \ge 3$, we need the summation
\[
M_j(n) = \sum_{s=3}^{n} \Bigl\{ \sum_{\gamma_{s}=0}^2 \sum_{\gamma_{s-1}=0}^2 \cdots \sum_{\gamma_3=0}^2 B_j\bigl(n,\vec{\gamma}_{n,s}\bigr) \Bigr\}  
\]
for the vertices $x_{\vec{\gamma}_{n,s}}$ with $s \ge 3$. By (\ref{RBr}) and (\ref{BBER}), $M_j(3) = \sum_{\gamma_3=0}^2 B_j(3,\vec{\gamma}_{3,3})$ is given by 
\beqn
M_j(3) & = & \Bigl\{ B_j \bigl(2,\vec{\gamma}^{0}_{2,2}\bigr) E_0 + B_j \bigl(2,\vec{\gamma}^{1}_{2,2}\bigr) E_1 +
B_j \bigl(2,\vec{\gamma}^{2}_{2,2}\bigr) E_2 \Bigr\} R \\
& = & B_j \bigl(2,(1,1)\bigr) [E_0+E_1+E_2] R \\
& = & B_j \bigl(1,0\bigr) R [E_0+E_1+E_2] R = B_j \bigl(1,0\bigr) RER \ ,
\eeqn
where
\[
E = E_0+E_1+E_2 = \left[\begin{array}{ccccc}
3 & 0 & 0 & 0 & 0 \\
0 & 2 & 1 & 0 & 0 \\
0 & 1 & 1 & 1 & 0 \\
0 & 0 & 1 & 2 & 0 \\
0 & 0 & 0 & 0 & 3
\end{array}
\right] \ .
\]
The general expression for $n\geq 3$ is 
\beqs
\lefteqn{M_j(n+1)} \cr\cr 
& = & \sum_{s=3}^{n+1} \Bigl\{ \sum_{\gamma_{s}=0}^2 \sum_{\gamma_{s-1}=0}^2 \cdots \sum_{\gamma_3=0}^2 B_j\bigl(n+1,\vec{\gamma}_{n+1,s}\bigr) \Bigr\} \cr
& = & \Bigl\{ B_j\bigl(n,\vec{\gamma}^{0}_{n,2}\bigr)E_0 + B_j\bigl(n,\vec{\gamma}^{1}_{n,2}\bigr)E_1 +
B_j\bigl(n,\vec{\gamma}^{2}_{n,2}\bigr)E_2 \Bigr\} R \cr
& & + \sum_{s=4}^{n+1} \Biggl\{ \sum_{\gamma_{s}=0}^2 \sum_{\gamma_{s-1}=0}^2 \cdots \sum_{\gamma_4=0}^2 \Bigl[ B_j\bigl(n,\vec{\gamma}^{0}_{n,s-1}\bigr)E_0 + B_j\bigl(n,\vec{\gamma}^{1}_{n,s-1}\bigr)E_1 + B_j\bigl(n,\vec{\gamma}^{2}_{n,s-1}\bigr)E_2 \Bigr] R \Biggr\} \cr
& = & \Bigl[ B_j\bigl(n,(n-1,1)\bigr) + \sum_{s=3}^{n} \Bigl\{ \sum_{\gamma_{s}=0}^2 \sum_{\gamma_{s-1}=0}^2 \cdots \sum_{\gamma_3=0}^2 B_j\bigl(n,\vec{\gamma}_{n,s}\bigr) \Bigr\} \Bigr] E R \cr
& = & \Bigl\{ B_j\bigl(n-1,n-2\bigr)R + M_j(n) \Bigr\} E R \cr
& = & B_j\bigl(n-1,n-2\bigr) R E R + B_j\bigl(n-2,n-3\bigr) R (ER)^2 + M_j(n-1) (ER)^2 \cr
& = & \cdots \cr
& = & \sum_{m=1}^{n-1} B_j\bigl(m,m-1\bigr) R (ER)^{n-m} \ .
\label{sum}
\eeqs
For example,
\beqn
\Phi_j(3) & = & 3F_j(3,o) + X_j(3,0) + X_j(3,1) + X_j(3,2) \\
& & + 2\bigl[ Y_j(3,1) + Y_j(3,2) + (1,1,1)B_j\bigl(1,0\bigr) R E R e_1 \bigr] \cr
& = & 3F_j(3,o) + (1,1,1) \Bigl\{ B_j(1,0)L^2 + B_j(2,1)L + B_j(3,2) \\
& & + 2\bigl[ B_j(1,0)RL + B_j(2,1)R + B_j(1,0) R E R \bigr] \Bigr\} e_1 \cr
& = & 3F_j(3,o) + (1,1,1) \Bigl\{ B_j(1,0) \bigl[L^2+2RL + 2 R E R \bigr] + B_j(2,1) \bigl[L + 2R\bigr] + B_j(3,2) \Bigr\} e_1 \ .
\eeqn
For general $n \geq 3$, we have
\beq
\Phi_j(n) = 3F_j(n,o) + \sum_{m=0}^{n-1} X_j(n,m) + 2\sum_{m=1}^{n-1} Y_j(n,m) + 2(1,1,1) \Bigl\{ \sum_{m=3}^{n} M_j(m) L^{n-m} \Bigr\} e_1 \ , 
\label{Phin}
\eeq
with $X_j(n,m)$ and $Y_j(n,m)$ given in (\ref{Xnm}) and (\ref{Ynm}), respectively. From (\ref{sum}), the summation in the last term of (\ref{Phin}) is
\beqn
\sum_{m=3}^{n} M_j(m)L^{n-m} & = & \sum_{m=3}^{n} \Bigl[ \sum_{s=1}^{m-2} B_j(s,s-1) R (ER)^{m-1-s} \Bigr] L^{n-m}\cr
& = & \sum_{m=2}^{n-1} \Bigl[ \sum_{s=1}^{m-1} B_j(s,s-1) R (ER)^{m-s} \Bigr] L^{n-1-m}\cr
& = & \sum_{s=1}^{n-2} B_j(s,s-1) \sum_{m=s+1}^{n-1} \Bigl[ R (ER)^{m-s} L^{n-1-m} \Bigr] \cr
& = & \sum_{s=1}^{n-2} B_j(s,s-1) \sum_{m=1}^{n-s-1} \Bigl[ R (ER)^{m} L^{n-1-m-s} \Bigr] \ ,
\eeqn
so that $\Phi_j(n)$ can be calculated exactly for any positive integer $n$. 

\begin{propo}
\label{propositionphi}
For the Sierpinski gasket $SG(n)$ with $n \geq 3$, the summation of $F_j(n,x)$ over all the vertex is given by
\beqs
\Phi_j(n) & = & (1,1,1) \Biggl\{ \sum_{m=1}^n B_j(m,m-1)R^{n-m} + 2\sum_{m=1}^{n-1} B_j\bigl(m,m-1\bigr) R L^{n-m-1} \cr
& & + 2\sum_{m=1}^{n-2} B_j(m,m-1) \sum_{s=1}^{n-m-1} \Bigl[ R (ER)^s L^{n-1-m-s} \Bigr] \Biggr\} e_1 + 3F_j(n,o) \cr
& = & (1,1,1) \Biggl\{ \sum_{m=1}^n B_j(m,m-1) R^{n-m} + 2B_j\bigl(n-1,n-2\bigr) R \cr
& & + 2\sum_{m=1}^{n-2} B_j\bigl(m,m-1\bigr) \Bigl[ \sum_{s=0}^{n-m-1} R (ER)^s L^{n-m-1-s} \Bigr] \Biggr\} e_1 + 3F_j(n,o) \ . 
\label{Phi}
\eeqs
\end{propo}

Let us consider limiting distribution 
\[
\lim_{n \to \infty} \phi_j(n) \equiv \phi_j
\]
with $j \in \{1,2,3,4\}$. It is easy to see that the term $3F_j(n,o)$ in (\ref{Phi}) can be neglected in the infinite $n$ limit for $\phi_j$, namely,
\[
\lim_{n\rightarrow\infty} \frac{F_j(n,o)}{\frac 32(3^n+1)} = 0 
\]
as the value $F_j(n,o)$ is between 0 and 1 for the four possible $j$. Similarly, the values of the quantities $F_j(m,x_{m-1})$, $G_j(m,x_{m-1})$, $H_j(m,x_{m-1})$ with $x=a,b,c$ in the matrix $B_j(m,m-1)$ are between 0 and 1, and all the eigenvalues of $R$ given in (\ref{R}) are positive and less than or equal to $1$, such that
\[
\lim_{n\rightarrow\infty} \frac{ (1,1,1) \Bigl\{ \sum_{m=1}^n B_j(m,m-1) R^{n-m} + 2B_j\bigl(n-1,n-2\bigr) R \Bigr\} e_1} {\frac 32(3^n+1)} \leq
\lim_{n\rightarrow\infty} \frac{3n+6}{\frac 32(3^n+1)} = 0 \ .
\]
Therefore, only the double summation term in (\ref{Phi}) gives non-zero contribution for $\phi_j$.
Rewrite $ER=Q_1[D_1+\bar{D}_1]Q_1^{-1}$ and $L=Q_2D_2Q_2^{-1}$, where
\[
Q_1 = \left[\begin{array}{ccccc}
159 & -87 & 0  & -3 & -3 \\
38  & 14  & 1  & 2  & 1  \\
38  & 14  & 0  & 2  & -4 \\
38  & 14  & -1 & 2  & 1  \\
15  & 45  & 0  & -3 & 5
\end{array}
\right] \ , \quad Q_2 = \left[\begin{array}{ccccc}
18 & 0  & -27 & 0  & -2 \\
5  & -1 & 98  & -1 & 1  \\
5  & 0  & -32 & 1  & 1  \\
0  & 1  & -52 & 0  & 0  \\
0  & 0  & 13  & 0  & 0
\end{array}
\right] \ ,
\]
and
\[
D_1 = \left[\begin{array}{ccccc}
0 & 0 & 0       & 0           & 0 \\
0 & 1 & 0       & 0           & 0 \\
0 & 0 & \frac25 & 0           & 0 \\
0 & 0 & 0       & \frac 3{25} & 0 \\
0 & 0 & 0       & 0           & 0
\end{array}
\right] \ , \quad \bar{D}_1 = \left[\begin{array}{ccccc}
3 & 0 & 0 & 0 & 0 \\
0 & 0 & 0 & 0 & 0 \\
0 & 0 & 0 & 0 & 0 \\
0 & 0 & 0 & 0 & 0 \\
0 & 0 & 0 & 0 & 0
\end{array}
\right] \ , \quad D_2 = \left[\begin{array}{ccccc}
1 & 0          & 0          & 0           &0 \\
0 & \frac3{10} & 0          & 0           &0 \\
0 & 0          & \frac6{25} & 0           &0 \\
0 & 0          & 0          & \frac{1}{5} & 0 \\
0 & 0          & 0          & 0           &\frac {1}{15}
\end{array}
\right] \ ,
\]
then
\beqs
\sum_{s=0}^{n-m-1} R (ER)^s L^{n-m-1-s} 
& = & R \sum_{s=0}^{n-m-1} \Bigl[ Q_1 (D_1^s + \bar{D}_1^s) Q_1^{-1} \Bigr] \Bigl[ Q_2 D_2^{n-m-1-s} Q_2^{-1} \Bigr] \cr
& = & R \sum_{s=0}^{n-m-1} \Bigl[ Q_1 D_1^s Q_1^{-1} \Bigr] \Bigl[ Q_2 D_2^{n-m-1-s} Q_2^{-1} \Bigr] \cr & &
+ R \sum_{s=0}^{n-m-1} \Bigl[ Q_1 \bar{D}_1^s Q_1^{-1} \Bigr] \Bigl[ Q_2 D_2^{n-m-1-s} Q_2^{-1} \Bigr] \ .
\label{D1D2}
\eeqs
The elements of $B_j(m,m-1)$ have been solved in (\ref{ivFn})-(\ref{ivHn}). Define 
\beqs
\lefteqn{Z_j(m) = (1,1,1) B_j(m,m-1)} \cr
& = & \left\{\begin{array}{l}
\lambda_1^{(1)} (\frac35)^m + \lambda_2^{(1)}(\frac1{25})^m + \lambda_3^{(1)} (\frac1{375})^m \ \mbox{for} \ j=1 \ , \cr
\frac{363}{196} \lambda_0 + \lambda_1^{(2)} (\frac35)^m + \lambda_2^{(2)} (\frac1{25})^m + \lambda_3^{(2)} (\frac1{375})^m + \lambda_4^{(2)} (\frac1{15})^m + \lambda_5^{(2)} (\frac1{225})^m \ \mbox{for} \ j=2 \ , \cr
\frac{99}{98}\lambda_0 + \lambda_1^{(3)}(\frac35)^m + \lambda_2^{(3)} (\frac1{25})^m + \lambda_3^{(3)} (\frac1{375})^m + \lambda_4^{(3)} (\frac1{15})^m + \lambda_5^{(3)} (\frac1{225})^m \ \mbox{for} \ j=3 \ , \cr
\frac{27}{196} \lambda_0 + \lambda_1^{(4)} (\frac35)^m + \lambda_2^{(4)} (\frac 1{25})^m + \lambda_3^{(4)} (\frac1{375})^m + \lambda_4^{(4)} (\frac 1{15})^m + \lambda_5^{(4)} (\frac1{225})^m \ \mbox{for} \ j=4 \ ,
\end{array}
\right. 
\label{Z}
\eeqs
where $\lambda_0=(1,1,1,1,1)$ and
\beqn
& & \lambda_1^{(1)}=(\frac{605}{392}, \frac{121}{56}, \frac{121}{56}, \frac{121}{56}, \frac{1089}{392}) \ , \
\lambda_2^{(1)} = (\frac{-1375}{196}, \frac{-55}{28}, \frac{-55}{28}, \frac{-55}{28}, \frac{1221}{196}) \ , \\
& & \lambda_3^{(1)} = (\frac{3125}{392}, \frac{-375}{56}, \frac{-375}{56}, \frac{-375}{56}, \frac{585}{392}) \ , \ 
\lambda_1^{(2)} = (\frac{-275}{396}, \frac{-55}{56}, \frac{-55}{56}, \frac{-55}{56}, \frac{-495}{392}) \ , \\
& & \lambda_2^{(2)} = (\frac{2375}{196}, \frac{95}{28}, \frac{95}{28}, \frac{95}{28}, \frac{-2109}{196}) \ , \
\lambda_3^{(2)} = (\frac{-9375}{392}, \frac{1125}{56}, \frac{1125}{56}, \frac{1125}{56}, \frac{-1755}{392}) \ , \\
& & \lambda_4^{(2)} = (\frac{-1265}{196}, \frac{-187}{196}, \frac{-187}{196}, \frac{-187}{196}, \frac{891}{196}) \ , \
\lambda_5^{(2)} = (\frac{10}{49}, \frac{-240}{49}, \frac{-240}{49}, \frac{-240}{49}, \frac {54}{49}) \ , \\
& & \lambda_1^{(3)} = (\frac{-285}{392}, \frac{-57}{56}, \frac{-57}{56}, \frac{-57}{56}, \frac{-513}{392}) \ , \
\lambda_2^{(3)} = (\frac{-625}{196}, \frac{-25}{28}, \frac{-25}{28}, \frac{-25}{28}, \frac{555}{196}) \ , \\
& & \lambda_3^{(3)} = (\frac{-9375}{392}, \frac{-1125}{56}, \frac{-1125}{56}, \frac{-1125}{56}, \frac{1755}{392}) \ , \
\lambda_4^{(3)} = (\frac{230}{49}, \frac{34}{49}, \frac{34}{49}, \frac{34}{49}, \frac{-162}{49}) \ , \\ 
& & \lambda_5^{(3)} = (\frac{500}{49}, \frac {480}{49}, \frac{480}{49}, \frac {480}{49}, \frac{-108}{49}) \ , \
\lambda_1^{(4)} = (\frac{-45}{392}, \frac{-9}{56}, \frac{-9}{56}, \frac{-9}{56}, \frac{-81}{392}) \ , \\ 
& & \lambda_2^{(4)} = (\frac{-375}{196}, \frac{-15}{28}, \frac{-15}{28}, \frac{-15}{28}, \frac{333}{196}) \ , \
\lambda_3^{(4)} = (\frac{-3125}{392}, \frac{375}{56}, \frac{375}{56}, \frac{375}{56}, \frac{-585}{392}) \ , \\ 
& & \lambda_4^{(4)} = (\frac{345}{196}, \frac{51}{196}, \frac{51}{196}, \frac{51}{196}, \frac{-243}{196}) \ , \quad 
\lambda_5^{(4)} = (\frac{250}{49}, \frac{-240}{49}, \frac{-240}{49}, \frac{-240}{49}, \frac{54}{49}) \ .
\eeqn
Substituting (\ref{D1D2}) into (\ref{Phi}), we get 
\beqs
\phi_j & = & \lim_{n\rightarrow\infty} \frac43 (3^n+1)^{-1} \sum_{m=1}^{n-2} Z_j(m)  R \Bigl\{ \sum_{s=0}^{n-m-1} \Bigl[ Q_1 D_1^s Q_1^{-1} \Bigr] \Bigl[ Q_2 D_2^{n-m-1-s} Q_2^{-1} \Bigr] \cr
& & + \sum_{s=0}^{n-m-1} \Bigl[ Q_1 \bar{D}_1^s Q_1^{-1} \Bigr] \Bigl[ Q_2 D_2^{n-m-1-s} Q_2^{-1} \Bigr] \Bigr\} e_1 \ .
\label{limphi}
\eeqs
As the eigenvalues of $D_1$ and $D_2$ are between 0 and 1, the first term in (\ref{limphi}) makes no contributions since
\beqn
0 & < & \sum_{m=1}^{n-2} Z_j(m)  R \sum_{s=0}^{n-m-1} \Bigl[ Q_1 D_1^s Q_1^{-1} \Bigr] \Bigl[ Q_2 D_2^{n-m-1-s} Q_2^{-1} \Bigr] e_1 \\
& \leq & \sum_{m=1}^{n-2} Z_j(m)  R \sum_{s=0}^{n-m-1} \Bigl[ Q_1 I Q_1^{-1} \Bigr] \Bigl[ Q_2 I Q_2^{-1} \Bigr] e_1 \leq 3n^2 \ .
\eeqn
Consider the second term in (\ref{limphi}),
\beqn
& & \sum_{m=1}^{n-2} Z_j(m) R \sum_{s=0}^{n-m-1} \Bigl[ Q_1 \bar{D}_1^s Q_1^{-1} \Bigr] \Bigl[ Q_2 D_2^{n-m-1-s} Q_2^{-1} \Bigr] \\
& = & \sum_{m=1}^{n-2} Z_j(m) R \Bigl[ Q_1 \tilde{D}_1 Q_1^{-1} \Bigr] \sum_{s=0}^{n-m-1} 3^s \Bigl[ Q_2 D_2^{n-m-1-s} Q_2^{-1} \Bigr] \\
& = & \sum_{m=1}^{n-2} Z_j(m) R \Bigl[ Q_1 \tilde{D}_1 Q_1^{-1} \Bigr] Q_2 D(n,m) Q_2^{-1} \ ,
\eeqn
where
\[
\tilde{D}_1 = \left[\begin{array}{ccccc}
1 & 0 & 0 & 0 & 0 \\
0 & 0 & 0 & 0 & 0 \\
0 & 0 & 0 & 0 & 0 \\
0 & 0 & 0 & 0 & 0 \\
0 & 0 & 0 & 0 & 0
\end{array}
\right]
\]
and define
\beqn
\lefteqn{D(n,m) \equiv \sum_{s=0}^{n-m-1} 3^s D_2^{n-m-1-s}} \cr
& = & \sum_{s=0}^{n-m-1} \left[\begin{array}{ccccc}
3^s (\frac 6{25})^{n-m-1-s} & 0 & 0 & 0 & 0 \\
0 & 3^s (\frac 1{15})^{n-m-1-s} & 0 & 0 & 0\\
0 & 0 & 3^s  & 0 & 0 \\
0 & 0 & 0 & 3^s (\frac {1}{5})^{n-m-1-s} & 0 \\
0 & 0 & 0 & 0 & 3^s (\frac {3}{10})^{n-m-1-s}
\end{array}
\right] \cr
& = & \left[\begin{array}{ccccc}
\frac{25[3^{n-m}-(\frac {6}{25})^{n-m}]}{69} & 0 & 0 & 0 & 0 \\
0 & \frac{15[3^{n-m}-(\frac 1{15})^{n-m}]}{44} & 0 & 0 & 0 \\
0 & 0 & \frac{3^{n-m}-1}{2} & 0 & 0 \\
0 & 0 & 0 & \frac{5[3^{n-m}-(\frac 15)^{n-m}]}{14} & 0 \\
0 & 0 & 0 & 0 & \frac{10[3^{n-m}-(\frac 3{10})^{n-m}]}{27}
\end{array}
\right] \cr\cr
& \equiv & 3^{n-m}{\mathcal D} + {\mathcal D}_2(n-m) \ ,
\eeqn
with
\[
{\mathcal D} = \left[\begin{array}{ccccc}
\frac{25}{69}& 0 & 0 & 0 & 0 \\
0 & \frac{15}{44} & 0 & 0 & 0 \\
0 & 0 & \frac{1}{2}  & 0 & 0 \\
0 & 0 & 0 & \frac{5}{14} & 0 \\
0 & 0 & 0 & 0 &\frac{10}{27}
\end{array}
\right] , 
\]
\[
{\mathcal D}_2(n-m) = \left[\begin{array}{ccccc}
\frac{-25(\frac {6}{25})^{n-m}}{69} & 0 & 0 & 0 &0 \\
0 & \frac{-(\frac 1{15})^{n-m-1}}{44} & 0 & 0 & 0 \\
0 & 0 & \frac{-1}{2} & 0 & 0 \\
0 & 0 & 0 & \frac{-(\frac 15)^{n-m-1}}{14} & 0 \\
0 & 0 & 0 & 0 & \frac{-(\frac 3{10})^{n-m-1}}{9}
\end{array} 
\right] \ .
\]
Since the absolute values of all the eigenvalues of ${\mathcal D}_2(n-m)$ are less than one, we have 
\beqs
\phi_j & = & \lim_{n\rightarrow\infty} \frac43 (3^n+1)^{-1} \sum_{m=1}^{n-2} \Bigl\{ 3^{n-m} Z_j(m) R Q_1 \tilde{D}_1 Q_1^{-1} Q_2{\mathcal D} Q_2^{-1} \cr
& & + Z_j(m) R Q_1 \tilde{D}_1 Q_1^{-1} Q_2 {\mathcal D}_2(n-m) Q_2^{-1} \Bigr\} e_1 \cr
& = & \lim_{n\rightarrow\infty}\frac43 (3^n+1)^{-1} \sum_{m=1}^{n-2} 3^{n-m} Z_j(m) \tilde R e_1 \ ,
\label{phij}
\eeqs
where $\tilde R = RQ_1\tilde{D}_1Q_1^{-1} Q_2{\mathcal D}Q_2^{-1}$.
Substituting the expression of $Z_j(m)$ from (\ref{Z}) into (\ref{phij}), carrying out the summation and taking the infinite $n$ limit, we arrive at
\beqn
\phi_j 
= \frac43 \times \left\{\begin{array}{ll}
& \Bigl [ \frac{\lambda_1^{(1)}}{4} + \frac{\lambda_2^{(1)}}{74} + \frac{\lambda_3^{(1)}}{1124}\Bigr ] \tilde R e_1 \quad \mbox{for} \quad j=1 \ , \cr
& \Bigl [ \frac{363\lambda_0}{392} + \frac{\lambda_1^{(2)}}{4} + \frac{\lambda_2^{(2)}}{74} + \frac{\lambda_3^{(2)}}{1124} + \frac{\lambda_4^{(2)}}{44} + \frac{\lambda_5^{(2)}}{674}\Bigr ] \tilde R e_1 \quad \mbox{for} \quad j=2 \ , \cr
& \Bigl [ \frac{99\lambda_0}{196} + \frac{\lambda_1^{(3)}}{4} + \frac{\lambda_2^{(3)}}{74} + \frac{\lambda_3^{(3)}}{1124} + \frac{\lambda_4^{(3)}}{44} + \frac{\lambda_5^{(3)}}{674}\Bigr ] \tilde R e_1 \quad \mbox{for} \quad j=3 \ , \cr
& \Bigl [ \frac{27\lambda_0}{392} + \frac{\lambda_1^{(4)}}{4} + \frac{\lambda_2^{(4)}}{74} + \frac{\lambda_3^{(4)}}{1124} + \frac{\lambda_4^{(4)}}{44} + \frac{\lambda_5^{(4)}}{674}\Bigr ] \tilde R e_1 \quad \mbox{for} \quad j=4 \ .
\end{array}
\right. \\
\eeqn
The matrix productions can be done to give the following theorem.

\begin{theo}
\label{theoremphi}
Consider all the vertices of the Sierpinski gasket $SG(n)$ in the infinite $n$ limit. The average probabilities that a vertex is connected by 1, 2, 3 or 4 bond(s) among all the spanning tree configurations are
\beqn
\phi_1 & = & \frac{10957}{40464} = 0.270783906682 \cdots \ , \qquad 
\phi_2 = \frac{6626035}{13636368} = 0.485909077842 \cdots \ , \cr\cr
\phi_3 & = & \frac{2943139}{13636368} = 0.215830124267 \cdots \ , \qquad 
\phi_4 = \frac{124895}{4545456} = 0.0274768912073 \cdots \ .
\eeqn
\end{theo}

\begin{cor}
Consider all the vertices of the Sierpinski gasket $SG(n)$ in the infinite $n$ limit. Denote $\theta$ as the average number of bonds connecting to a vertex among all the spanning tree configurations, then
\[
\theta = \phi_1 + 2\phi_2 + 3\phi_3 + 4\phi_4 = 2 \ .
\]
\end{cor}

We list the numerical values of $\phi_j(n)$ with $j \in \{1,2,3,4\}$ for $0 \le n \le 5$ and infinite $n$ limit in Table \ref{phitable}. We find that $\phi_1(n)$ decreases monotonically as $n$ increases, while $\phi_2(n)$, $\phi_3(n)$ and $\phi_4(n)$ increase monotonically. The values for $n=5$ are already very close to $\phi_j$ in the infinite $n$ limit with deviations about $1\%$. 

\begin{table}
\caption{Numerical values of $\phi_j(n)$ with $j \in \{1,2,3,4\}$, and the comparison of $\phi_j$ with $f_j$ for the square lattice ($sq$). The last digits given are rounded off.}
\begin{center}
\begin{tabular}{|c|c|c|c|c|}
\hline\hline 
$n$ & $\phi_1(n)$ & $\phi_2(n)$ & $\phi_3(n)$ & $\phi_4(n)$ \\ \hline\hline 
0 & $\frac23$     & $\frac13$       &   &   \\ 
  & =0.6666666667 & =0.3333333333   & 0 & 0 \\ \hline
1 & $\frac12$ & $\frac{19}{54}$ & $\frac{7}{54}$  &  $\frac1{54}$  \\ 
  & =0.5      & =0.3518518519   & =0.1296296296   & =0.01851851852 \\ \hline 
2 & $\frac{163}{450}$ & $\frac{5257}{12150}$ & $\frac{2203}{12150}$ & $\frac{289}{12150}$ \\
  & =0.3622222222 & =0.4326748971 & =0.1813168724 & =0.02378600823 \\ \hline
3 & $\frac{143357}{472500}$ & $\frac{17871899}{38272500}$ &  $\frac{7787951}{38272500}$  & $\frac{1000733}{38272500}$ \\ 
  & =0.3034010582 & =0.4669645045 & =0.2034868639 & =0.02614757332 \\ \hline 
4 & $\frac{24381607}{86484375}$ & $\frac{30227565716}{63047109375}$ &  $\frac{13341669059}{63047109375}$  & $\frac{1703683097}{63047109375}$ \\ 
  & =0.2819192137 & =0.4794441175 & =0.2116142864 & =0.02702238237 \\ \hline 
5 & $\frac{39739246273}{144755859375}$  & $\frac{51047283737324}{105527021484375}$ & $\frac{22626394285676}{105527021484375}$  & $\frac{2883432928358}{105527021484375}$ \\ 
  & =0.2745259946 & =0.4837366110  & =0.2144132751 & =0.02732411934 \\ \hline 
$\infty$ & $\frac{10957}{40464}$ & $\frac{6626035}{13636368}$ & $\frac{2943139}{13636368}$ & $\frac{124895}{4545456}$ \\ 
  & =0.2707839067 & =0.4859090778 & =0.2158301243 & =0.02747689121 \\ 
\hline\hline 
$sq$ & $f_1$ & $f_2$ & $f_3$ & $f_4$ \\ \hline
  & $\frac{8}{\pi^2}-\frac{16}{\pi^3}$ & $\frac{8}{\pi}-\frac{36}{\pi^2}+\frac{48}{\pi^3}$ & $2-\frac{16}{\pi}+\frac{48}{\pi^2}-\frac{48}{\pi^3}$ & $-1+\frac{8}{\pi}-\frac{20}{\pi^2}+\frac{16}{\pi^3}$ \\ 
  & =0.2945449182 & =0.4469901311 & =0.2223849831 & =0.03607996755 \\ \hline\hline
\end{tabular}
\end{center}
\label{phitable} 
\end{table}

It is interesting to compare the Sierpinski gasket $SG(n)$ in the infinite $n$ limit with the infinite two-dimensional square lattice which is also a 4-regular lattice. For the square lattice, all the vertices are identical due to the translational invariant. The probabilities that a vertex is connected by 1, 2, 3 or 4 bond(s) among all the spanning tree configurations have been solved exactly in \cite{mdm} that were denoted as $f_j$ with $j \in \{1,2,3,4\}$. As shown in Table \ref{phitable}, $f_1$, $f_3$ and $f_4$ are slightly larger than $\phi_1$, $\phi_3$, $\phi_4$, respectively, while $f_2$ is smaller than $\phi_2$. Especially, the average number of bonds connecting to a vertex among all the spanning tree configurations on the square lattice $f_1+2f_2+3f_3+4f_4$ is equal to two, which is exactly the same as $\theta$ here for the Sierpinski gasket.

Acknowledgement:
The research of S.C.C. was partially supported by the NSC
grant NSC-96-2112-M-006-001. The research of L.C.C was partially supported by TJ$\&$MY Foundation and the NSC grant NSC-96-2115-M-030-002. L.C.C. would like to thank PIMS, university of British Columbia for the hospitality. We would like to thank Akira Sakai for valuable comments.

\vfill
\eject
\end{document}